\documentclass[useAMS,usenatbib]{mnras}
\usepackage{amsmath}
\usepackage{appendix}
\usepackage{graphicx}
\usepackage{xfrac}
\usepackage{caption}
\usepackage{array}
\usepackage{pdflscape}
\usepackage[figuresright]{rotating}
\usepackage{booktabs}
\usepackage[usenames, dvipsnames]{color}
\usepackage{subcaption}
\usepackage[flushleft]{threeparttable}
\usepackage[export]{adjustbox}
\usepackage{hyperref}
\allowdisplaybreaks

\usepackage{etoolbox}

\title[GAMA: forensic SED reconstruction ]{Galaxy And Mass Assembly (GAMA): A forensic SED reconstruction of the cosmic star formation history and metallicity evolution by galaxy type}
\author[Bellstedt et al.]
{Sabine Bellstedt,$^{1}$\thanks{Email: sabine.bellstedt@uwa.edu.au} Aaron S. G. Robotham,$^{1, 2}$ Simon P. Driver,$^{1, 3}$ Jessica E. Thorne,$^{1}$ 
	\newauthor Luke J. M. Davies,$^{1}$ Claudia del P. Lagos,$^{1, 2}$ Adam R. H. Stevens,$^{1, 2}$ 
	\newauthor Edward N. Taylor,$^{4}$ Ivan K. Baldry,$^{5}$ Amanda J. Moffett,$^{6}$ Andrew M. Hopkins,$^{7}$ 
	\newauthor Steven Phillipps$^{8}$
\\
$^{1}$ ICRAR, The University of Western Australia, 7 Fairway, Crawley WA 6009, Australia\\
$^{2}$ ARC Centre of Excellence for All Sky Astrophysics in 3 Dimensions (ASTRO 3D) \\
$^{3}$ SUPA, School of Physics \& Astronomy, University of St Andrews, North Haugh, St Andrews, KY16 9SS, UK \\
$^{4}$ Centre for Astrophysics and Supercomputing, Swinburne University of Technology, Hawthorn, VIC 3122, Australia \\
$^{5}$ Astrophysics Research Institute, Liverpool John Moores University, IC2, Liverpool Science Park, 146 Brownlow Hill, Liverpool, L3 5RF\\
$^{6}$ Department of Physics and Astronomy, University of North Georgia, 3820 Mundy Mill Rd., Oakwood GA 30566, USA \\
$^{7}$ Australian Astronomical Optics, Macquarie University, 105 Delhi Rd, North Ryde, NSW 2113, Australia\\
$^{8}$ Astrophysics Group, School of Physics, University of Bristol, Tyndall Avenue, Bristol BS8 1TL, UK \\
}

\begin{document}

\date{}

\pagerange{\pageref{firstpage}--\pageref{lastpage}} \pubyear{2020}

\maketitle

\label{firstpage}

\begin{abstract}

We apply the spectral energy distribution (SED) fitting code \textsc{ProSpect} to multiwavelength imaging for $\sim$7,000 galaxies from the GAMA survey at $z<0.06$, in order to extract their star formation histories. 
We combine a parametric description of the star formation history with a closed-box evolution of metallicity where the present-day gas-phase metallicity of the galaxy is a free parameter. 
We show with this approach that we are able to recover the observationally determined cosmic star formation history (CSFH), an indication that stars are being formed in the correct epoch of the Universe, on average, for the manner in which we are conducting SED fitting. 
We also show the contribution to the CSFH of galaxies of different present-day visual morphologies, and stellar masses. 
Our analysis suggests that half of the mass in present-day elliptical galaxies was in place 11 Gyr ago. In other morphological types, the stellar mass formed later, up to 6 Gyr ago for present-day irregular galaxies. 
Similarly, the most massive galaxies in our sample were shown to have formed half their stellar mass by 11 Gyr ago, whereas the least massive galaxies reached this stage as late as 4 Gyr ago (the well-known effect of ``galaxy downsizing").
Finally, our metallicity approach allows us to follow the average evolution in gas-phase metallicity for populations of galaxies, and extract the evolution of the cosmic metal mass density in stars and in gas, producing results in broad agreement with independent, higher-redshift observations of metal densities in the Universe. 

\end{abstract}


\section{Introduction}


A basic test of our understanding of galaxy formation and evolution in a cosmological context is to derive a consistent description of the cosmic star formation history (CSFH) with the evolution of the galaxy stellar mass function. 

The CSFH, which describes the total star formation rate across all galaxies per unit comoving volume as a function of time, has predominantly been constructed by measuring the instantaneous star formation rates of galaxies over a wide redshift range in a ``core-sample" approach \citep[as has been done in numerous studies, for example][]{Madau96, Hopkins00, Giavalisco04, Ouchi04, Hopkins04, Thompson06, Hopkins06, Verma07, Karim11, Robotham11, Cucciati12, Sobral13, RowanRobinson16, Davies16, Wang19}, and recently to $z \sim 5$ by \citet{Driver18} using a compilation of the GAMA \citep{Driver11, Liske15}, G10-COSMOS \citep{Davies15, Andrews17} and 3DHST \citep{Momcheva16} surveys. 
These studies show that star formation peaks around 3.5 Gyr after the Big Bang and that the cosmic star formation rate has declined exponentially since. There is, however, still significant debate as to the exact position of the CSFH peak \citep[as discussed in detail in sec. 6 of][]{Hopkins18}, with highly dust-obscured systems at high redshift possibly missed \citep[e.g. ][]{TWang19}, potentially causing an underestimation of the true peak redshift. 

A related property to the CSFH is the stellar mass density (SMD). 
The SMD describes the total amount of stellar mass present in the Universe as a function of cosmic time. This is observationally derived by integrating under the galaxy stellar mass function at different epochs. Such an analysis has been frequently conducted, by studies including \citet{Caputi11, Caputi15, Gonzalez11, Mortlock11, Mortlock15, Santini12, Muzzin13, Duncan14, Tomczak14, Grazian15, Song16, Davidzon17, Driver18}. The important thing to note is that, for any given object, the evolving stellar mass can be inferred from the star formation history (SFH) through integration while taking into account mass lost via mechanisms such as stellar winds and supernovae, and hence the SMD can be derived from the CSFH. Observationally derived CSFH and SMD curves are, however, often shown to be inconsistent \citep[as discussed by][]{Wilkins08, Wilkins19, Hopkins18}, highlighting the presence of underlying unknowns that are affecting the successful extraction of the CSFH and SMD (such as potential variations in the IMF, or observational brightness limits). 
Recent work by \citet{Davidzon18} has shown that it is possible to measure specific star formation rates of high-$z$ galaxies based on the differential evolution of the SMD, highlighting that in the future we may be able to invert the process described above to derive the CSFH from the SMD. 

A complementary but contrasting method of recovering the CSFH and SMD is to extract the individual SFHs of low-redshift galaxies that are encoded in their spectral energy distributions (SEDs). Techniques that apply this method are reliant on stellar population models \citep[for example][]{Silva98, Charlot00, Bruzual03, Maraston05, Conroy09}. These produce typical spectra for stellar populations of varying ages that can be fitted against data to identify the fractions of light produced by stars of particular ages. 
This can either be done by fitting to a spectrum, as done in the codes STECMAP \citep{Ocvirk06}, Vespa \citep{Tojeiro07}, and STARLIGHT \citep{CidFernandes11}; or by fitting to an SED consisting of broadband photometry, as done by \textsc{MAGPHYS} \citep{daCunha08}, CIGALE \citep{Noll09}, \textsc{Prospector} \citep{Johnson17, Leja17} and BAGPIPES \citet{Carnall19}. The former has the advantage of higher spectral resolution, enabling features such as absorption lines to be fitted, whereas the latter has the advantage of a wide wavelength range, enabling a fit from the far-ultraviolet (FUV) to far-infrared (FIR) simultaneously.
We note that a disadvantage of the former is that spectra frequently suffer from aperture effects, unlike photometric data.
Some codes, like CIGALE, BAGPIPES and \textsc{Prospector}, are able to simultaneously fit both broadband SEDs and spectra, ensuring that the benefits of both approaches can be utilised. 

The advantage of a ``forensic" technique like this is that, by construction, the resulting CSFH and SMD will be consistent, and hence the evolution of SFR and stellar mass can be studied simultaneously. 
While forensic techniques to measure the CSFH have been present in the literature for many years \citep[with earlier examples including][]{Heavens04, Panter07}, there has been a recent resurgence in their popularity \citep[including ][]{Leja18, Carnall19, LopezFernandez18, Sanchez19}.
These studies have used a mix of photometric, spectroscopic and integral field unit (IFU) data in order to extract SFHs for individual galaxies, and have had varying degrees of success in recovering the directly observed CSFH. In particular, recovering a consistent position of the peak in the CSFH has been elusive. 

One of the major differences between different spectral-fitting codes is the manner in which the star formation histories are parametrized. Generally, the codes come in two flavours: parametric, and non-parametric. Parametric methods describe the star formation histories by an analytic function (like the exponentially declining model used in \textsc{MagPhys}), whereas non-parametric methods (which would be better described as uncontinuous functional methods, as they are still described parametrically) tend to fit with different age bins, allowing for a discontinuous and somewhat arbitrary SFH shape for individual galaxies. 
The merits of parametric versus non-parametric methods have been debated in the literature \citep[for example][]{Carnall19, Leja18}, with both methods commonly employed. Parametric fits are generally physically motivated, and have fewer parameters to be fitted. Non-parametric fits inevitably have more free parameters, but significantly more flexibility as to the types of SFH produced. Care needs to be taken, however, to ensure that ``unphysical" star formation histories are not produced, and that the constraining power of the data on the free parameters is sufficient. 

While the particular parametrization of SFHs has received much attention in the literature, the implementation of metallicity evolution has been less explored, with most studies, assuming an unphysical history where the metallicity is constant with time for simplicity. 
This is perhaps surprising, as the concept of the age--metallicity degeneracy has been known for a long time \citep[e.g.][]{Worthey94, Dorman03, Cole09}, with the implication that age and metallicity impact an observed SED in similar ways. 
Significant effort has been invested over previous decades to measure the age and metallicities of stellar populations using both spectral data \citep[e.g.][]{Tang09, Woodley10, Gallazzi14, Feltzing17} and photometric data \citep[e.g.][]{Piatti11, Tang13, Piatti14}, accompanied by theoretical work \citep{Romero15}, in order to overcome the limitations introduced by this degeneracy. 
The retrieved distribution of ages (i.e., the star formation history), will inevitably be significantly impacted by the assumed evolution in metallicity, and as such a physical treatment of metallicity evolution \citep[as explored by][]{Driver13} is crucial to improve the quality of SED-fitting outputs, as we will demonstrate in this paper. 

In this work, we apply the SED-fitting code \textsc{ProSpect}\footnote{Available at \url{https://github.com/asgr/ProSpect}} \citep{Robotham20} in a parametric mode to multiwavelength photometry from the GAMA survey in order to measure the star formation histories of $\sim$7,000 galaxies at $z<0.06$.
\textsc{ProSpect} is advantageous for this purpose due to its flexible and modular nature, allowing not only the SFH to be parametrised in any way, but also the evolution of the gas-phase metallicity. 
We will show that, in combination with a physically motivated implementation of the metallicity evolution, this approach successfully replicates the observational CSFH.  
The successful extraction of the CSFH is an essential first step in the retrieval of individual galaxy parameters, and provides a pathway to study the histories of galaxy populations as a function of environment or destination morphology. 

The structure of the paper is as follows. Our data are outlined in Sec. \ref{sec:Data}, with our adopted method described in Sec. \ref{sec:Method}. The results of our analysis are shown in Sec. \ref{sec:SFRDandSMD} and Sec. \ref{sec:Metallicity}, followed by a discussion of our caveats in Sec. \ref{sec:Discussion}. We summarise our results in Sec. \ref{sec:Summary}. 
We present in Appendix \ref{sec:ModelComparison} a brief comparison of our results to simulations, and Appendix \ref{sec:LinearMetallicity} presents the main results with an alternate implementation of evolving metallicity. 
All stellar masses quoted in this paper are calculated assuming a \citet{Chabrier03} IMF, and all magnitudes are in the AB system. 
The cosmology assumed throughout this paper is $H_0 = 67.8\,\rm{km}\,\rm{s}^{-1}\,\rm{Mpc}^{-1}$,  $\Omega_m = 0.308$ and $\Omega_{\Lambda} = 0.692$ \citep[consistent with a Planck 15 cosmology:][]{Planck16}.


\section{Data}
\label{sec:Data}

We use the spectroscopic and photometric data from the GAMA survey \citep{Driver11, Liske15}. This survey was a large spectroscopic campaign on the Anglo Australian Telescope that gathered redshifts for $\sim$300,000 galaxies in five fields (G02, G09, G12, G15 and G23), amounting to a total sky area of 230 square degrees. GAMA targets were selected by size and colour above a magnitude limit of $m_r \leq 19.8$ (or $m_i \leq 19.0$ in G23), and the survey achieved a high spectroscopic completeness of 98 per cent to the magnitude limits in the equatorial regions, in order to successfully conduct environmental science.

We use the far-UV -- far-IR photometry derived using  the source-finding software \textsc{ProFound}\footnote{Available at \url{https://github.com/asgr/ProFound}} \citep{Robotham18}, as described by \citet{Bellstedt20}, from the \texttt{GAMAKidsVikingFIRv01} Data Management Unit (DMU).
The photometric bands from this data release include: GALEX $FUV$ and $NUV$; VST $u,g,r,i$; VISTA $Z, Y,J,H,K_S$; WISE $W1$, $W2$, $W3$, $W4$; and Herschel $P100$, $P160$, $S250$, $S350$ and $S500$ \citep[see][for more details on the data genesis]{Driver16}. 
The photometric extraction is outlined in detail by \citet{Bellstedt20}, but in brief, \textsc{ProFound} is applied on an $r+Z$ band stack for the initial source detection, and then in multiband mode, where it is applied to the full optical $FUV-W2$ bands. For a subset of the optically-extracted photometry expected to be detectable in the FIR, \textsc{ProFound} is used in \texttt{FitMagPSF} mode in order to obtain fluxes in the $W3-S500$ bands, which are semi- to unresolved. 

We exclusively use galaxies with a redshift quality flag $NQ > 2$, i.e. where spectroscopic redshifts are reasonably certain ($P>90\%$). For this work, we restrict the redshift range of the sample to $z<0.06$, producing a volume-limited sample. 
This redshift range is selected to be large enough to encompass sufficiently many galaxies, but small enough to be targeting only galaxies within the last 0.8 Gyr of cosmic time.
Additionally, the galaxies within this redshift range have visually classified morphologies available, providing a desirable avenue for further analysis. 
In addition, we select only those objects that have been classified as galaxies in the photometric catalogue, as given by \textsc{uberclass}$=$\texttt{galaxy}. 
Objects in the photometric catalogue are assigned a \texttt{galaxy} class on the basis of both size and colour \citep[although see][for a detailed discussion of the star--galaxy separation applied]{Bellstedt20}.
Based on the updated GAMA photometry presented in \citet{Bellstedt20}, we remeasure the 95\% completeness limit in the $r$-band (the selection limits) to be 19.5/19.5/19.5/19.0 in the G09/G12/G15/G23 fields respectively. 
We hence restrict our sample to $m_r \leq 19.5$ in only the equatorial fields, G09, G12 and G15 to ensure uniform completeness throughout the sample.

\begin{figure}
	\centering
	\includegraphics[width=85mm]{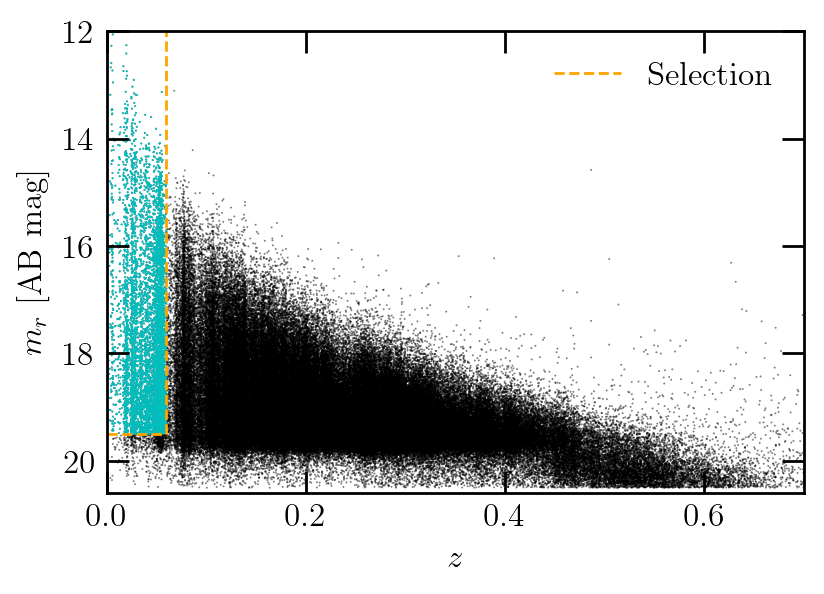}
	\caption{Galaxy redshift sample from the GAMA G09/G12/G15 fields (black) and the subset of objects used in this work shown in cyan, where $m_r$ values are taken from \citet{Bellstedt20}. The orange dashed line shows the selection cut used, given by $z<0.06$ and $m_r \leq 19.5$. }
	\label{fig:SampleSelection}
\end{figure}

Based on the survey area measurements stated in \citet{Bellstedt20}, we implement a survey area of 169.3 square degrees for the three combined GAMA equatorial regions. 
Hence the sample contains 6,688 galaxies with $z<0.06$ and $m_r \leq 19.5$ in the G09/G12/G15 fields. We show this sample selection in Fig. \ref{fig:SampleSelection}. 

These visual morphologies used in this study come from the \texttt{VisualMorphologyv03} DMU \citep{Moffett16a}. 
In our analysis, we separate galaxies into four classes: elliptical galaxies (\texttt{E}); \texttt{S0-Sa} galaxies (including both \texttt{S0-Sa} and \texttt{SB0-SBa} classifications); spiral galaxies (including both \texttt{Sab-Scd} and \texttt{SBab-SBcd} classifications); and finally \texttt{Sd-Irr} galaxies.

\section{Method}
\label{sec:Method}


\subsection{\textsc{ProSpect}}

\textsc{ProSpect} \citep{Robotham20} is an energy balance SED-fitting code. Much like other SED-fitting codes, it makes use of stellar libraries to create an unattenuated stellar SED, which is then attenuated by dust, and subsequently re-emitted by the dust in the far-IR. 
A key feature of \textsc{ProSpect} is its modularity. Modules such as the selected parametrization of the star formation history (either in a parametric or a non-parametric form), or the metallicity evolution can be entirely user-defined. 
For the analysis presented in this work, we select the \citet{Bruzual03} stellar population models suite, as it has the appropriate wavelength range for the GAMA data, for SED-fitting. 

We note that using an energy-balance approach requires the assumption that the absorption of UV light produced by young stars and emission of IR light by dust is exactly balanced. Any spatial variation in these features has the potential to disrupt this balance, as geometric features in the dust could over- or under-attribute the amount of UV light that is being attenuated. Previous studies using UV and optical broadband imaging and spatially-resolved analysis show that the integrated, photometric approach can underestimate stellar masses inferred by spatially resolved techniques due to the presence of dust lanes \citep{Zibetti09}. In this work, we do not directly address this general limitation of energy-balance SED modelling.   

In the following sections, we outline the parametrizations selected for both the star formation history, and the metallicity evolution.

\subsubsection{SFH parametrization}

\textsc{ProSpect} has a number of inbuilt parametric and non-parametric SFH models, as described in detail by \citet{Robotham20}. 
The parametrization of the SFH that has been selected for use in this paper is the \texttt{massfunc\_snorm\_trunc}, which is a skewed normal distribution that is anchored to zero star formation at the beginning of the Universe. 
The parametrization of the \texttt{snorm} SFH is given as follows:
\begin{equation}
\rm{SFR}(t)_{\rm snorm}=\texttt{mSFR}\times e^{\frac{-X(t)^2}{2}},
\end{equation}
where
\begin{equation}
X(t) = \left(\frac{t-\texttt{mpeak}}{\texttt{mperiod}}\right) {\left(e^{\texttt{mskew}}\right)}^{\rm{asinh}\left(\frac{t-\texttt{mpeak}}{\texttt{mperiod}}\right)} 
\end{equation}
This parametrization has four free parameters:
\begin{itemize}
	\item{\texttt{mSFR} -- the peak SFR of the SFH,}
	\item{\texttt{mpeak} -- the age of the SFH peak,}
	\item{\texttt{mperiod} - the width of the normal distribution, }
	\item{\texttt{mskew} -- the skewness of the normal distribution. }
\end{itemize}
To ensure that the SFH is zero at the start of the Universe, we implement a truncation over the above parametrization. This is conducted as follows:

\begin{align}
\rm{SFR}(t)_{\rm trunc}&= \rm{SFR}(t)_{\rm snorm} \times \\ 
& \left[1 - \frac{1}{2} \left[1+\rm{erf}\left( \frac{t-\mu}{\sigma\sqrt{2}}\right)   \right]\right] \nonumber
\end{align}
where
\begin{align}
\mu&= \texttt{mpeak} +\frac{|(\texttt{magemax} - \texttt{mpeak})|} {\texttt{mtrunc}}\\ 
\sigma&=\frac{|(\texttt{magemax} - \texttt{mpeak})|} {2\times\texttt{mtrunc}}
\end{align}
 This parametrization achieves a smooth truncation (with no discontinuities) between the peak of the SFH and the beginning of the Universe. For this work, we use a fixed value of $\texttt{mtrunc}=2$ Gyr and $\texttt{magemax}=13.4$ Gyr. Note that this truncation implementation introduces no additional free parameters in our analysis. 

Examples of star formation histories derived using this parametrization have been shown in fig. 10 of \citet{Robotham20}. 
The forced truncation in the early Universe of this parametrization is favourable, as it provides a strong constraint on the possible shape of the SFH in an epoch that is poorly constrained from the SED itself. 
The implicit assumption made in selecting this parametrization is that the star formation rates are rising in the first billion years of the Universe. 
We select the \texttt{magemax} parameter to fix the start of star formation to the epoch at which the highest-$z$ galaxies are known to exist ($z=11$, \citealt{Oesch16}), corresponding to a lookback time of 13.4 Gyr. We highlight, however, that other work on a $z=9$ galaxy suggests that it was forming stars as early as $z=15$ \citep{Hashimoto18}, and hence this \texttt{magemax} value could be regarded as a lower limit.


The \texttt{massfunc\_snorm\_trunc} parametrization is inherently unimodal, and will achieve the best results for galaxies that have experienced a single epoch of star formation. 
For galaxies that may experience two distinct periods of star formation, this parametrization will be inaccurate. 
However as shown in \citet{Robotham20}, based on a comparison of \textsc{ProSpect} fits to galaxies from the semi-analtyic model \textsc{Shark} \citep{Lagos19}, this parametrization of the SFH is able to recover the SFH of a population of galaxies fairly accurately (see Sec \ref{sec:Discussion} for a further discussion of caveats). 
As such, while the resulting SFHs may be a poor description of the true SFH for some individual galaxies, we expect to derive reasonable histories for galaxy populations.

\subsubsection{Metallicity parametrization}
\label{sec:MetallicityParametrization}
In most typical SED-fitting implementations, metallicity is fixed to a constant value throughout the history of a galaxy, with the exact value of this fixed metallicity generally allowed to be a free parameter \citep[for example, see the works of][]{Leja18, Carnall19}. 

The approach that we have taken in this work is to evolve the metallicity at the same rate as the build-up of stellar mass, assuming a closed-box metallicity evolution, as given by the \texttt{Zfunc\_massmap\_box} parametrization within \textsc{ProSpect}.  
Within the closed-box model, each galaxy only has a fixed amount of gas available with which to form stars. Throughout the history of the galaxy, gas is converted into stars according to its SFR. Over time, this gas is enriched via a fixed yield, which specifies the fraction of metal mass produced in stars that is returned to the gas.  
While a closed-box metallicity evolution will likely be unrealistic for galaxies with large gas inflows/outflows or for galaxies interacting strongly with their environment, we expect that this assumption will produce reasonable results at a statistical level. Additionally, this approach will provide a significant improvement over the assuption of a constant metallicity over cosmic time. 
In this parametrization of the metallicity, the starting and ending metallicities can be set (either as fixed values, or as free parameters), while the shape is determined by the derived star formation history of an individual galaxy with the assumption of closed-box stellar evolution. 
Examples of metallicity histories derived using this parametrization can be seen in fig. 13 of \citet{Robotham20}. 
The final metallicity of the galaxy ($Z_{\rm final}$) is treated as a free parameter in our approach, and is allowed to range between $10^{-4}$ and $5\times10^{-2}$ (corresponding to the metallicity range of the \citet{Bruzual03} templates). We fix the initial metallicity to $10^{-4}$ for each galaxy, which corresponds to the minimum metallicity of the \citet{Bruzual03} templates. 

Finally, as discussed in \citet{Robotham20}, the metallicity yield in this approach (the ratio of metal mass released into the ISM to the mass locked up in stars), is a parameter that can be varied within \textsc{ProSpect}. In our implementation, we adopt a fixed value of 0.02 for all galaxies. 
This value is slightly smaller than values typically implemented in the literature, For example, a value of 0.03 was implemented by \citealt{Peeples14} having explored values in the range 0.0214--0.0408, the semi-analytic model \textsc{Shark} assumes a value of 0.029 \citealt{Lagos19}, and similarly the semi-analytic model \textsc{Dark SAGE} \citep{Stevens18} uses a value of 0.03\footnote{For the semi-analytic models, this yield value was selected based on a \citet{Chabrier03} IMF with a \citet{Conroy09} simple stellar population.}. In reality, yield values decline during a galaxy's lifetime (the yields from superovae reduce as the metallicity increases, see for example \citealt{Kobayashi06}), and therefore assuming a constant yield of 0.03 would result in an overestimation of the metal mass of galaxies at late stages of their evolution. The slightly lower value implemented in this study results in more realistic metal masses throughout a galaxy's history. 
We note, however, that the impact of changing the yield in the range 0.02-0.03 on our CSFH is negligible.

As will be reflected in this work, the incorporation of a closed-box metallicity evolution represents a signifcant advance over the commonly adopted free-but-constant metallicity in SED-fitting codes (demonstrated in Sec. \ref{sec:MetallicityImpact}).

\subsection{\textsc{ProSpect} fitting}
\label{sec:ProSpectFitting}

\subsubsection{Data setup}
\label{sec:DataSetup}

In the wavelength range $6-50\,\mu m$, \textsc{ProSpect} model SEDs are dominated by polyaromatic hydrocarbon (PAH) features produced by dust, which are highly susceptible to modelling assumptions. The $W3$ photometric band exists within this region, and as a result of these features, we find that the measured fluxes are poorly modelled by \textsc{ProSpect}. We find that by including the $W3$ photometry measurements, we are biasing our SED fits, and are not able to adequately model the FIR peak. To avoid this, we have opted to exclude all $W3$ measurements from fitting. 

We show in Table \ref{tab:PhotometrySummary} the fraction of objects that are missing photometric measurements in the UV or IR bands, due to a lack of coverage and due to non-detections. We do not include non-detections in our fitting, as we find that they artificially suppress the fitted FIR peak. In our sample, 20 per cent of the galaxies are missing all Herschel data (P100-S500), with 60 per cent of those objects missing all Herschel data due to a lack of FIR coverage. We find that galaxies with no FIR detections tend to have slightly lower dust masses, but observe no biases for objects missing data due to lack of FIR coverage. As expected, the constraint on the resulting dust mass and SFR reduces if the FIR data are missing.  

\begin{table}
	\centering
	\caption[PhotometrySummary]{Summary of photometric details for our sample, including the percentage of objects for which we have no data in each band, the percentage of objects for which no flux was detected, and the error floor added in each band to the uncertainty measurements to account for modelling variations. }
	\label{tab:PhotometrySummary}
	\begin{tabular}{@{}c | ccc}
		\hline
		Band&  Missing & Non-detected & Error Floor \\
		& (\%) & (\%) & \\
		(1)& (2) &(3) & (4) )\\
		\hline
		FUV  & 8 & 0 & 0.16 \\
		NUV  & 4 & 0 &0.16 \\
		$u$  & 0 & 0 &  0.1 \\
		$g$  & 0 & 0 & 0.04 \\
		$r$  & 0 & 0 &   0.03 \\
		$i$ & 0 & 0 &   0.045 \\
		$Z$  & 0 & 0 &  0.03\\
		$Y$  & 0 & 0 & 0.035 \\
		$J$  & 0 & 0 &  0.045 \\
		$H$  & 0 & 0 &  0.07 \\
		$K_S$  & 0 & 0 &   0.08 \\
		W1  & 0 & 0 &  0.05\\
		W2  & 0 &0  &  0.14 \\
		W4  & 0 & 45 & 0.165 \\
		P100 & 17 & 36 &  0.1 \\
		P160 & 17 & 32 &  0.1 \\
		S250 & 20 & 25 &  0.1 \\
		S350 & 19 & 30 &  0.1 \\
		S500 &19 & 39 &  0.1 \\
		\hline		
	\end{tabular}
\end{table}

We estimate the photometric variation that ensues due to modelling variations from the standard deviation of the photometric residuals. We use this measurement as a floor to augment the flux uncertainty value for each band, where the final error is given by $\sigma_{\rm final} = \sqrt{\sigma_{\rm obs}^2 + (\rm{floor}\times\rm{flux}_{\rm obs})^2}$. This floor value is shown in column 4 of Table \ref{tab:PhotometrySummary}. 

\subsubsection{MCMC setup}
\label{sec:MCMC}

We implement \textsc{ProSpect} in a Bayesian manner using Markov Chain Monte Carlo (MCMC) via the CHARM\footnote{Componentwise Hit-And-Run} algorithm provided by the \texttt{LaplacesDemon} package within R \citep{LaplacesDemon18}. MCMC robustly explores large and complex parameter spaces with potentially multi-modal solutions, and where there are strong covariances or degeneracies between parameters. 
MCMC is run using 10,000 steps for each galaxy, where the fitted parameters have generally been burned in after around 1000 steps. 
This burn-in is generally small, as a genetic algorithm\footnote{We implement a Covariance Matrix Adapting Evolutionary Strategy (CMA), as taken from \url{https://github.com/cran/cmaes}. } has been implemented to conduct a quick fit of the SED, the results of which are used as the initial parameter guesses for MCMC. 
With this configuration, \textsc{ProSpect} takes just over half an hour to run for a single galaxy on a single, standard processor. To process all $\sim7,000$ galaxies, over a 24-hour period, we used 28 cores on each of 6 nodes on the Zeus supercomputer at the Pawsey Supercomputing Centre. 

The filter response curves used for each of the photometric bands are taken from \textsc{ProSpect} \citep{Robotham20}, and are consistent with those used by \citet{Driver18} for their \textsc{MAGPHYS} analysis. 

\textsc{ProSpect} has the flexibility to allow each parameter to be fitted in either linear or logarithmic space, depending on which is more suitable for the dynamic range of each parameter (as specified by the \texttt{logged} input parameter). The parameter ranges and whether linear or logarithmic spaces were used are indicated in Table \ref{tab:MCMCSpecs}. 

Weak priors have also been imposed for the dust radiation field alpha parameters, in the form of a Normal distribution. Where relevant, the prior has been indicated in Table \ref{tab:MCMCSpecs}. These priors have been based on the distribution of best-fitting parameters when \textsc{ProSpect} is run without priors, as a method of constraining the parameters for galaxies in which the data are generally uninformative.  

An invariant \citet{Chabrier03} IMF has been assumed throughout our analysis, and the maximum age of galaxies (the lookback time at which star formation may begin) is set to 13.4 Gyr. 

\begin{table*}
	\centering
	\caption[MCMC Intervals]{Technical specifications for the MCMC implementation for the \texttt{massfunc\_snorm\_trunc} SFH parametrization. }
	\label{tab:MCMCSpecs}
	\begin{tabular}{@{}c | cccc |  c}
		\hline
		Parameter & \multicolumn{4}{c|}{MCMC configuration} &Units\\
		& Fitting & Logarithmic & Range & Prior & \\
		\hline
		\multicolumn{6}{l}{\textit{SFH parameters}} \\[3pt]
		\hline
		mSFR & Fitted & Yes & [$-3$, 4] &--& $\rm M_{\odot} {\rm yr}^{-1}$ \\
		mpeak & Fitted & No & [$-(2+t_{\rm lb})$, $13.4-t_{\rm lb}$]  &--& Gyr\\
		mperiod & Fitted & Yes & [$\log_{10}(0.3)$, 2] &--& Gyr  \\
		mskew & Fitted & No & [$-0.5$, 1] &--& -- \\
		\hline
		\multicolumn{6}{l}{\textit{Metallicity parameters}} \\[3pt]
		\hline
		Zfinal & Fitted & Yes & [$-4$, $-1.3$]& --& -- \\
		\hline
		\multicolumn{6}{l}{\textit{Dust parameters}} \\[3pt]
		\hline
		$\tau_{\rm birth}$ & Fitted & Yes & [$-2.5$, 1.5]& -- & -- \\
		$\tau_{\rm screen}$ & Fitted & Yes & [$-2.5$, 1] & --& --\\
		$\alpha_{\rm birth}$ & Fitted & No & [0, 4] & $\rm exp({-\frac{1}{2}(\frac{\alpha_{\rm birth}-2}{1})^2})$& --\\
		$\alpha_{\rm screen}$ & Fitted & No & [0, 4]  &$\rm exp({-\frac{1}{2}(\frac{\alpha_{\rm screen}-2}{1})^2})$& --\\
		${\rm pow}_{\rm birth}$ & Fixed & -- &--& -- &--  \\
		${\rm pow}_{\rm screen}$ & Fixed & --&-- & -- &--  \\
		\hline
	\end{tabular}
\end{table*}

\subsubsection{Example outputs}

\begin{figure*}
	\centering
	\includegraphics[width=180mm]{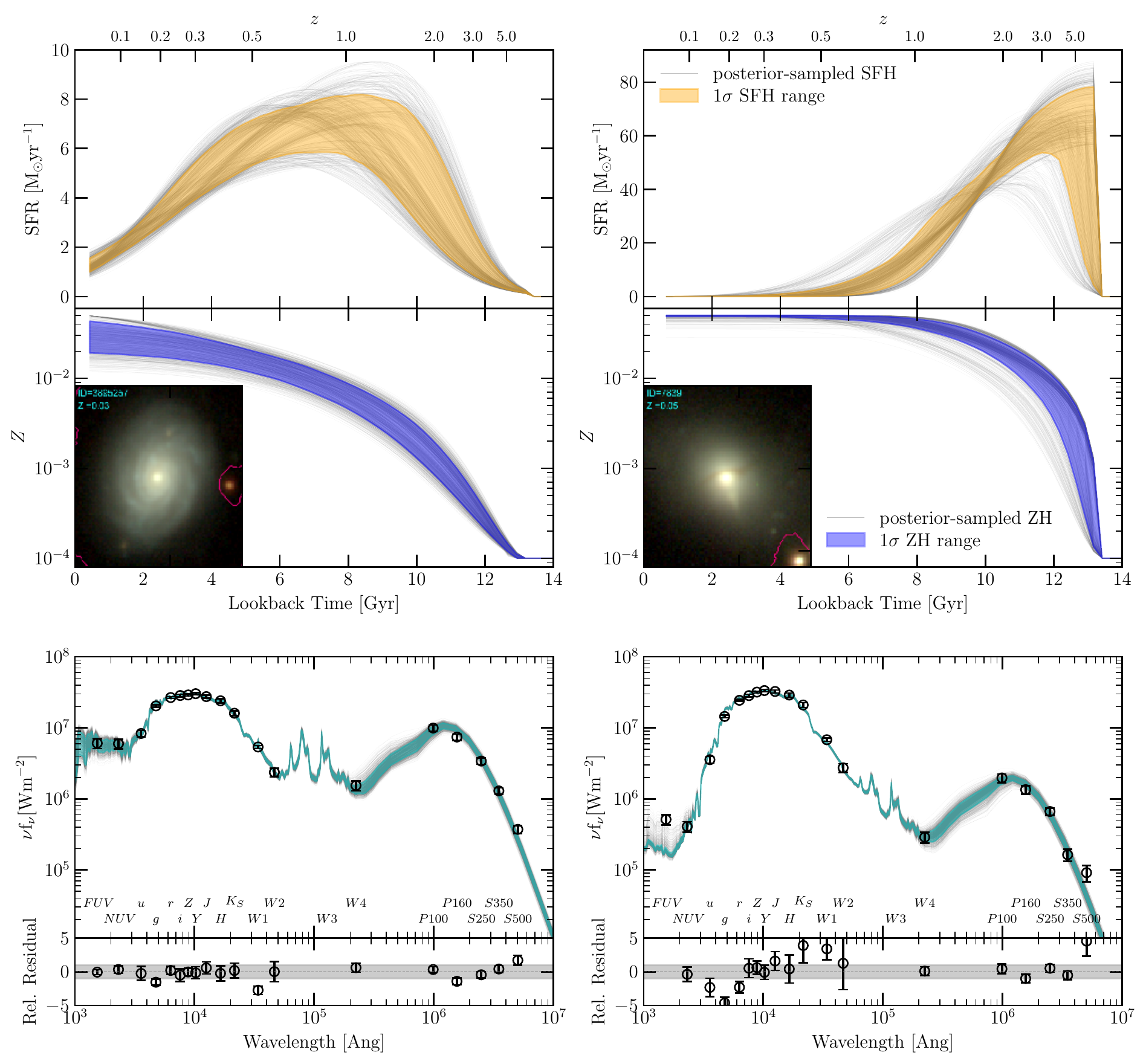}
	\caption{\textsc{ProSpect} outputs for the late-type galaxy 3895257 (left), and for the early-type galaxy 7839 (right). Top: Example star formation history. The grey lines are 1,000 star formation histories from the thinned MCMC posterior distribution, with the orange shaded region showing the 1$\sigma$ range of the SFH. Middle: Corresponding metallicity history, with the grey lines showing 1,000 MCMC histories, and the shaded blue region showing the 1$\sigma$ range of the metallicity history. A false-colour image of the galaxy is shown in the bottom left of this panel. Bottom: ProSpect SED fit to the observed SED (SEDs from the thinned posterior distribution are shown in grey, and the 1$\sigma$ range of the SED shown in cyan), with the residual fit shown below.  }
	\label{fig:SFH_examples}
\end{figure*}

Example outputs for two GAMA galaxies CATAID=3895257 and 7839 (respectively a late- and early-type) as derived by our implementation of ProSpect are shown in Fig. \ref{fig:SFH_examples}. 
The derived star formation histories are shown in the top panels.
Here, the orange shaded region shows the 1$\sigma$ range on the SFH, based on a thinned sample of SFHs from the posterior distribution, shown in grey. 
The 1$\sigma$ range on the corresponding metallicity evolution of each of the galaxies is shown in the middle panel in blue, again with the sampled metallicity histories shown in grey. 
From this panel it is clear to see how the build-up of metals follows the cumulative star formation.
The fits to the SEDs, including residuals, are shown in the lower sets of axes in Fig. \ref{fig:SFH_examples}.

The resulting variation in the star formation histories, shown by the spread of grey lines in Fig. \ref{fig:SFH_examples}, can be significant. The late-type galaxy peaked in star formation $\sim$6--10 Gyr ago, but is still forming stars at the present day, whereas the early-type galaxy has a star formation history for which star formation peaked $>$10 Gyr ago and stopped forming stars $\sim$ 6 Gyr ago. For the late-type galaxy (3895257), the prolonged star formation influences the resulting metallicity evolution, where the metallicity continues to increase until the present time. For the quenched early-type galaxy (7839), the maximum metallicity\footnote{Note that the maximum metallicity reached by this early-type galaxy is $5\times10^{-2}$, which corresponds to the maximum metallicity available by the \citet{Bruzual03} templates.} was reached at the time of quenching, with the metallicity being maintained from then on. \newline 

\begin{figure}
	\centering
	\includegraphics[width=85mm]{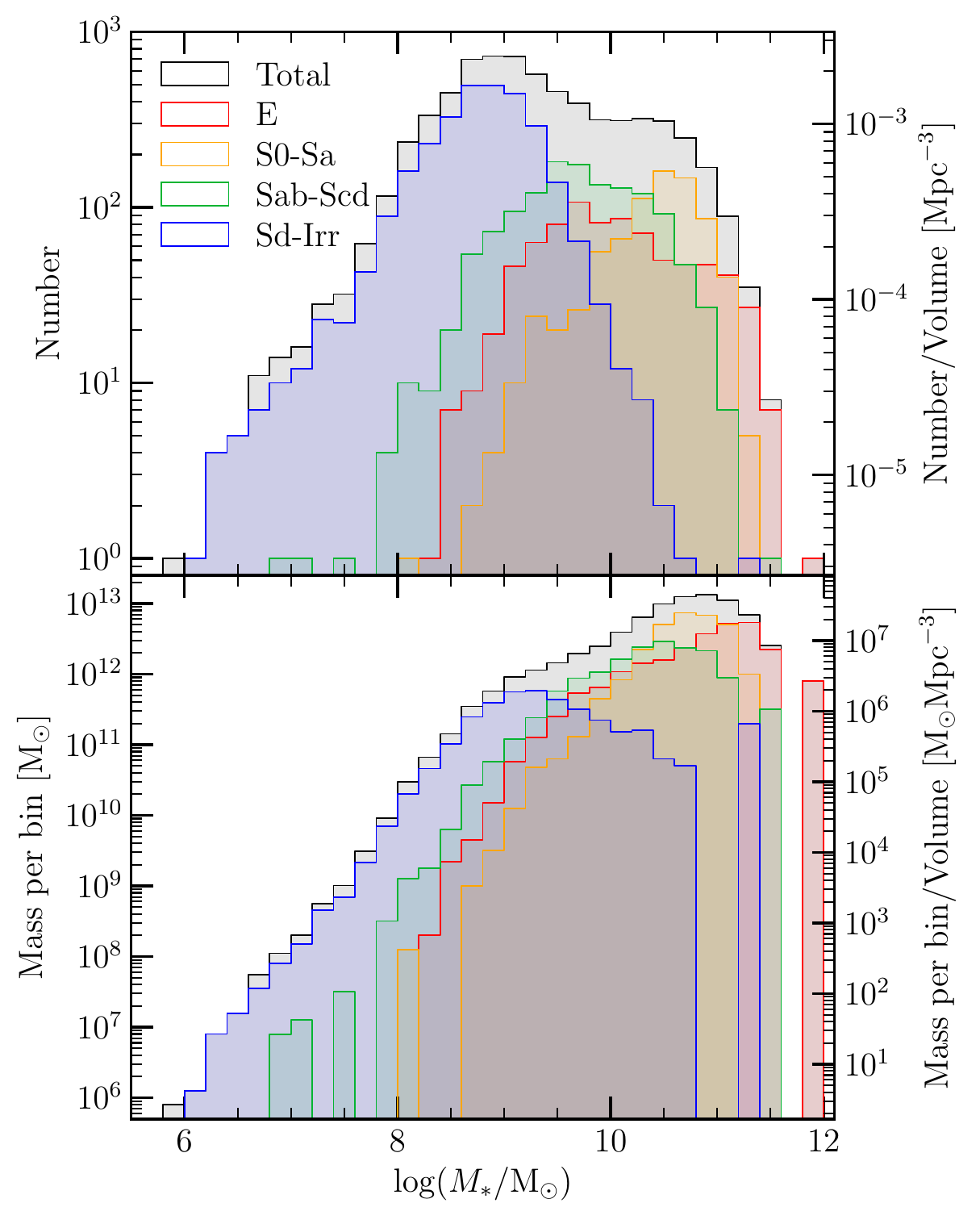}
	\caption{Top panel: Stellar mass distribution of the sample. In addition to the total distribution, we also show the distributions of stellar mass as determined by the individual morphological classifications. The number of objects per unit volume is indicated with the right-hand label. Lower panel: Total stellar mass present in each stellar mass bin. Each bin is 0.2 dex wide. }
	\label{fig:StellarMassHist}
\end{figure}

The resulting distribution of stellar masses of the galaxies in our sample is shown in the top panel of Fig. \ref{fig:StellarMassHist} for the full sample in grey. We also include the distribution of the galaxies as divided by their visual morphological classifications. These distributions agree with the expectation that the most massive galaxies are early-type galaxies, whereas the least massive are dominated by late-type galaxies. The lower panel of Fig. \ref{fig:StellarMassHist} shows the total amount of stellar mass present in each stellar mass bin. This panel highlights that most of the stellar mass in our sample comes from galaxies with $M_* > 10^{10} \rm{M}_{\odot}$, with irregular galaxies making only a very small contribution to the stellar mass of the sample.

\section{SFR/stellar mass densities}
\label{sec:SFRDandSMD}

\subsection{Measurement}
\label{sec:CSFHmeasurement}


In this work, we determine the CSFH by stacking the SFHs for each of the galaxies in our sample, normalised by the volume of the sample. We do this for each of the SFHs sampled from the MCMC distribution (shown in Fig. \ref{fig:SFH_examples} as the grey curves) in order to convey the sampling uncertainty  on the CSFH. 

To account for the mass incompleteness below stellar masses of $10^{9}\, \rm M_{\odot}$, we apply a stellar mass correction. To do this, we assume that the true mass distribution within our sampled volume can be described by the fitted double Schechter function from \citet{Kelvin14}. Dividing our sample into stellar mass bins of $\Delta \log(M_*/\rm M_{\odot}) = 0.1$, we calculate a ``correction factor" for each of the bins in our mass incomplete range. Then, the contribution towards the star formation rate and stellar mass densities from each stellar mass bin is multiplied by the derived correction factor. 
While we conduct this correction down to stellar masses of $\sim10^6\,\rm{M}_{\odot}$, (it is not possible to correct to lower stellar masses as there are no galaxies below this limit in our sample, although at this mass scale the contribution to the CSFH would be negligible), we note that by a stellar masses of $<10^8\, \rm{M}_{\odot}$ there are fewer than 100 galaxies per bin, and hence the stellar mass completeness correction is liable for inaccuracy. 
This correction makes a negligible change to the cosmic stellar mass density, but it noticeably increases the derived cosmic star formation rates at lookback times $< 4$ Gyr, as this is when low-mass galaxies experience most of their star formation (as we will show in Sec. \ref{sec:StellarMassTrends}). 

\subsubsection{Impact of metallicity on the CSFH}
\label{sec:MetallicityImpact}

\begin{figure}
	\centering
	\includegraphics[width=85mm]{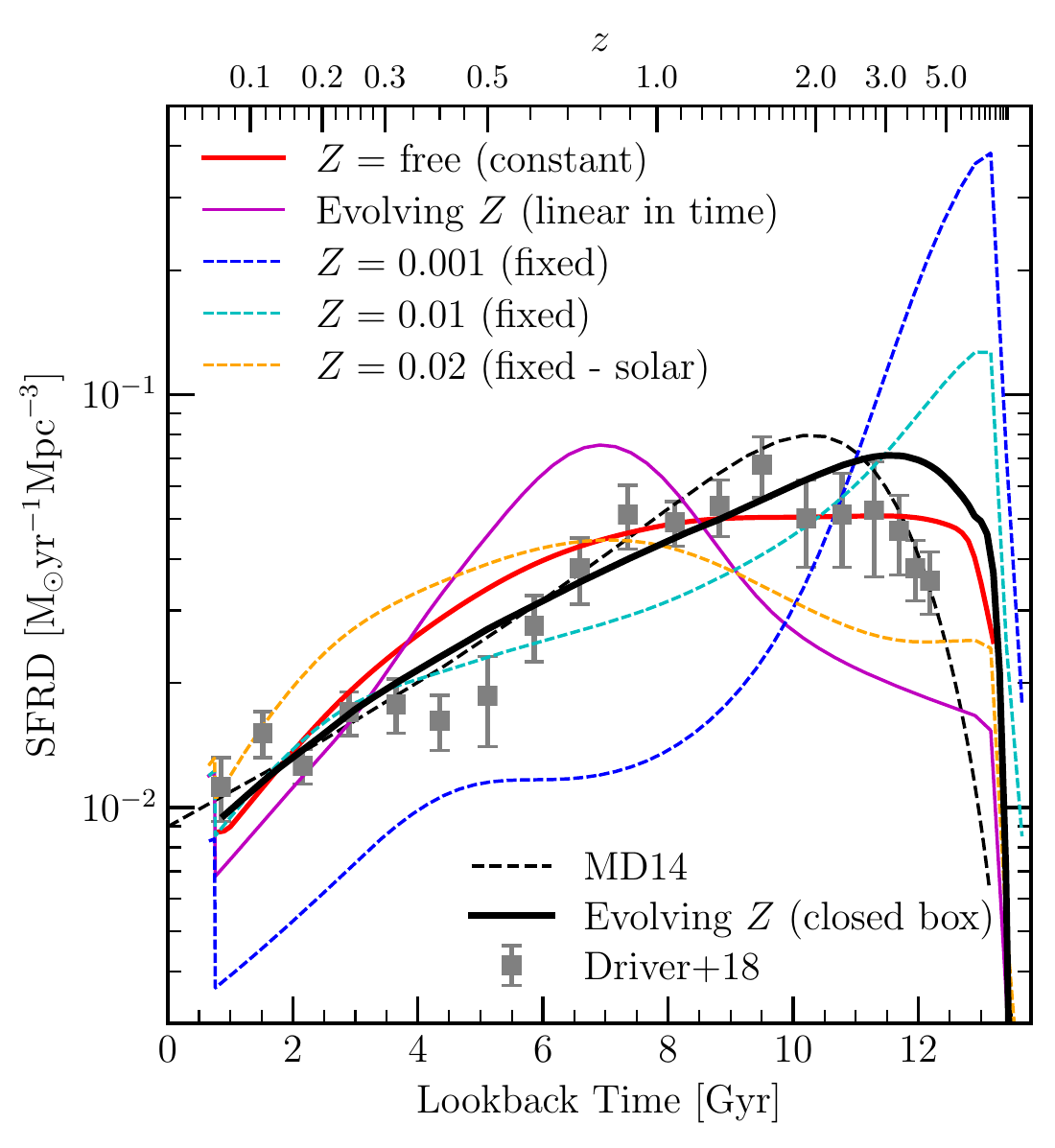}
	\caption{Effect of different metallicity assumptions on the resulting cosmic star formation history, using the \texttt{massfunc\_snorm\_burst} parametrization of the star formation history, as compared against the observational measurements by \citet{Driver18}, and the fit to the compilation of measurements by \citet{Madau14}. We show the CSFH derived in this work using a closed-box metallicity evolution in black.   }
	\label{fig:MetallicityComparison}
\end{figure}

The adopted metallicity evolution within SED-fitting techniques has an enormous impact on the accuracy of the SFH determination, and hence the ability to recover the correct evolution of the star formation density over cosmic time. 
By using the sample of $\sim$7,000 $z<0.06$ GAMA galaxies, we conduct a test by running \textsc{ProSpect} on each galaxy SED using different metallicity evolution assumptions, and constructing the corresponding CSFHs by stacking the resulting SFHs, as described in Sec. \ref{sec:CSFHmeasurement}.  The resulting CSFH curves from this test are shown in Fig. \ref{fig:MetallicityComparison}. 

We highlight that the SFH function adopted in this test is different to that in the main body of this paper: here we use \texttt{massfunc\_snorm\_burst} (with a star-forming burst as a free parameter, and no forced truncation of the SFH at the beginning of the Universe) rather than \texttt{massfunc\_snorm\_trunc}. 
In implementing this parametrization, the SFH is not anchored to 0 at the early Universe by the forced truncation of \texttt{massfunc\_snorm\_trunc}, and likewise the burst allows for flexibility in the late Universe for individual galaxies, while still constraining the most recent star formation. While this is not the ideal method of generating a CSFH (as we do not expect the CSFH to have a burst), this parametrization emphasises the effect of metallicity, and hence is useful in this illustration. 

Fig. \ref{fig:MetallicityComparison} shows the range in the CSFH that can be derived when the assumptions of metallicity evolution are altered. 
For reference, we include in Fig. \ref{fig:MetallicityComparison} the observational measurements by \citet{Driver18}, the fit to the compilation of measurements\footnote{Measurements included in this compilation come from the following studies: \citet{Sanders03, Takeuchi03, Wyder05, Schiminovich05, Dahlen07, Reddy09, Robotham11, Magnelli11, Magnelli13, Cucciati12, Bouwens12b, Bouwens12a, Schenker13, Gruppioni13}} by \citet{Madau14}, and the CSFH derived in this work.
For illustrative purposes, we depict very extreme assumptions of metallicity evolution. 
The most extreme is the CSFH derived when assuming that each galaxy had a constant metallicity of $Z=0.001$, shown in blue. 
In order to fit the observed SED, \textsc{ProSpect} is forced to produce very old stellar populations in order to combat the highly underestimated metallicity. 
This is evident in the resulting CSFH, for which the peak has been shifted to 13 Gyr, and all star formation in the past 10 Gyr has been severely underestimated. 
By increasing the metallicity by a factor of 10, the resulting CSFH with constant $Z=0.01$ is much closer to that observed, but the peak in star formation is still too old (cyan line). 
If the metallicity is increased significantly to solar metallicity, the assumption of constant $Z=0.02$ tends to be an overestimate of the true metallicity, and this is reflected in the corresponding CSFH (orange) where the peak in star formation is underestimated. 
Typically, the exact value of the metallicity is treated as a nuisance parameter in SED-fitting codes, and is allowed to be free \citep[as done by, for example,][]{Leja18, Carnall19}. 
The corresponding CSFH we derive when making this assumption is shown in Fig. \ref{fig:MetallicityComparison} in red. 
While the agreement with the observed CSFH is much better than in the previous extreme assumptions, we note that the clear peak seen in observational measurements is not reflected, showing that the assumption of constant metallicity washes out this feature. 
This highlights the necessity of allowing for metallicity evolution in individual galaxies when measuring their star formation histories. An example of a naive evolution is shown in Fig. 
\ref{fig:MetallicityComparison} in magenta, where the metallicity is assumed to be linearly evolving between 0 and solar throughout cosmic time. The consequence of such a naive approach is clear, with a peak now appearing much later, at $\sim 6$ Gyr. 

Note that with the \texttt{massfunc\_snorm\_trunc} SFH parametrization, this effect would be less dramatic, as the low-$z$ SFR is generally well-constrained, and therefore the most recent portion of the CSFH would be in much better agreement without the presence of a burst. Similarly, the forced truncation in the early Universe would somewhat suppress the dramatic rise in the CSFH seen in the lowest-metallicity demonstrations.

In summary, Fig. \ref{fig:MetallicityComparison} shows us that the frequently used assumption of a constant-but-free metallicity (red curve) can produce a similar shape to the empirical CSFH but for a washed out peak. A peak can be much better recovered using a naive linear metallicity evolution (magenta curve), but such an ill-motivated evolution recovers the peak at a potentially arbitrary position.

The above illustration shows that a physically motivated implementation of metallicity evolution is critical, in order to accurately recover the star formation histories of galaxies. This was also shown in section 4.1 and fig. 3 of \citet{Driver13}, where the implementation of an evolving metallicity linked to the CSFH was best able to reproduce the cosmic spectral energy distribution (cSED) for bulges and discs. 
We therefore use the closed-box metallicity evolution described in Sec. \ref{sec:MetallicityParametrization} throughout the analysis presented in this paper.

\subsection{Derived CSFH and SMD}

\begin{figure*}
	\centering
	\includegraphics[width=180mm]{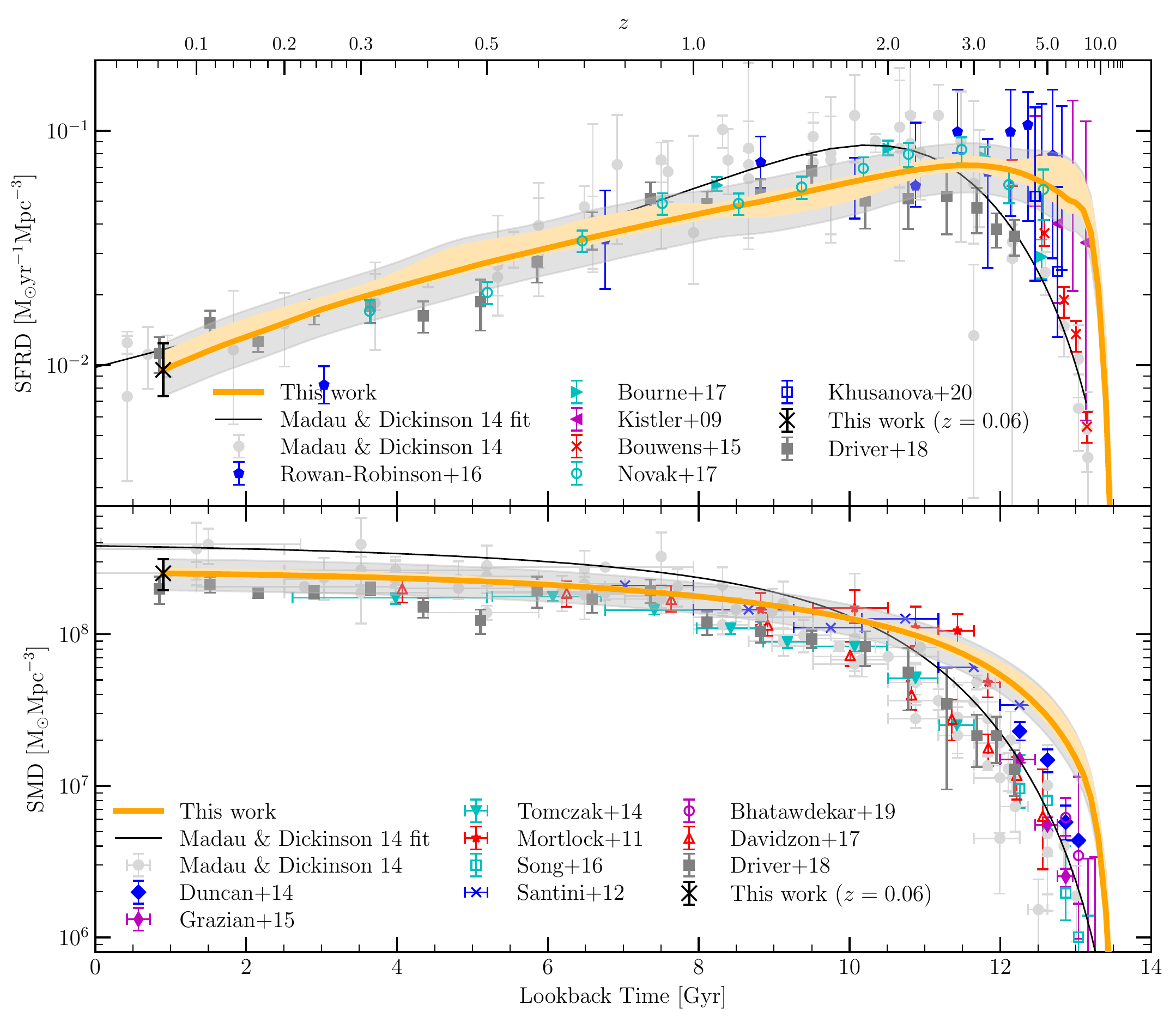}
	\caption{The cosmic star formation history (top panel) and stellar mass density (bottom panel) from \textsc{ProSpect} (shown in orange), compared with literature SFRD values. The top panel compares our CSFH with other core-sample results, including \citet{Kistler09, Madau14, Bouwens15, RowanRobinson16, Bourne17, Novak17, Driver18}. Clearly, the range in observed SFRD values at high-$z$ is still high. The bottom panel compares our cosmic SMD with core-sample observations, including \citet{Mortlock11, Santini12, Duncan14, Tomczak14, Grazian15, Song16, Davidzon17, Bhatawdekar19, Khusanova20}.   }
	\label{fig:cSFRComparisonLiterature}
\end{figure*}

The final \textsc{ProSpect}-derived CSFH and SMD are shown in the top and bottom panels of Fig. \ref{fig:cSFRComparisonLiterature} respectively.  Fig. \ref{fig:cSFRComparisonLiterature} shows that the CSFH recovered from a forensic analysis of $\sim7,000$ galaxies matches the core sample study of 600,000 galaxies across all redshifts by \citet{Driver18} well. The match is especially close at lookback times below 10 Gyr. 
While our result, shown in Fig. \ref{fig:cSFRComparisonLiterature} by the orange solid line, is largely consistent with the core-sampled CSFH of the last $\sim$ 10 Gyr, the derived peak in the CSFH occurs $\sim$ 1 Gyr earlier than in the measurements of \citet{Driver18}. 
The larger disagreement at earlier lookback times is a potential indication that individual galaxies in \textsc{ProSpect} experience an early increase in star formation that is too rapid. While the implemented truncation in our SFH parametrization ensures a tapering of the SFH at early times, galaxies that have a peak in star formation near the beginning of the Universe will be driving this result. Through implementation of a prior on the \texttt{mpeak} value that disfavours the peak occuring earlier than 12 Gyr it is possible to forcibly delay the peak in the CSFH. However we have elected not to include such a prior in this work. 
We highlight that there is still significant uncertainty in accounting for dust in observations of high-$z$ galaxies in studies such as that by \citeauthor{Driver18} \citep[as shown by studies such as][]{Reddy09, Gruppioni13, RowanRobinson16, Koprowski17, Novak17, Wang19}, and it is therefore difficult to determine the exact cause of disagreement at high redshift between forensic and core-sample methods. 

We indicate an estimate of the uncertainty due to MCMC sampling of SFH in Fig. \ref{fig:cSFRComparisonLiterature} (solid orange shaded region). 
As expected, the uncertainties are higher at earlier times as compared with the recent CSFH, highlighting that recent star formation is in general better constrained than older star formation. 
In addition to the sampling uncertainty, we indicate in the grey transparent shaded region the additional uncertainty that results from the cosmic variance of our $z<0.06$ sample, which is 22.8 per cent.\footnote{As derived using the ICRAR cosmology calculator \url{https://cosmocalc.icrar.org/}} 
Note that despite the potentially large uncertainties of the SFHs for individual galaxies, the form of the CSFH is robust to sampling from these varying SFHs. 

The SFRD measurement at $z<0.06$ is directly derived (rather than forensically), and as such is equivalent to other ``core-sample" measurements. We measure this value to be $(9.6^{+2.8}_{-2.2})\times 10^{-3}\, \rm{M}_{\odot}\,\rm{yr}^{-1}\,\rm{Mpc}^{-3}$, which we indicate in Fig. \ref{fig:cSFRComparisonLiterature} as a black cross. The equivalent measurement for the SMD is $(2.52^{+0.61}_{-0.58})\times10^8\, \rm{M}_{\odot}\,\rm{Mpc}^{-3}$. 

We additionally compare our CSFH against SFRD values as derived from gamma-ray bursts \citep{Kistler09}, values derived from the far-IR \citep{RowanRobinson16, Bourne17}, rest-frame UV \citep{Bouwens15}, a combination of far-IR and UV \citep{Khusanova20}, and from the radio \citep{Novak17}. We highlight that core-sample results derived using gamma-ray bursts, far-IR and radio techniques all result in SFRD values at high $z$ greater than those of \citet{Madau14}, more consistent with the SFRD derived by our analysis at this epoch.  
Whilst our CSFH is most influenced by modelling assumptions in the early epochs of the Universe (see Appendix \ref{sec:LinearMetallicity} for a discussion of how a slightly different implementation of metallicity evolution can impact the resulting CSFH and cSMD), our CSFH suggests a potential underestimation of the CSFH peak by the \citet{Madau14} fit. However, the cSMD that we recover is systematically higher than observations in the first 2 billion years of the Universe, suggesting that our peak may still be slightly overestimated. 

The cSMD shown in the bottom panel of Fig. \ref{fig:cSFRComparisonLiterature} also shows good agreement with the observed values of \citet{Driver18}. As a consequence of our CSFH being higher than the observations at lookback times greater than 11 Gyr, the corresponding cSMD is overestimated in the early Universe. 
We note that the stellar mass and SFR measurements by \citet{Driver18} were derived using \textsc{MAGPHYS}. 
As determined by \citet{Robotham20}, stellar masses derived by \textsc{ProSpect} are systematically 0.17 dex greater than those estimated by \textsc{MAGPHYS}. This is most likely because the metallicity implementation allows older stars to be formed, and hence more stellar mass must be recovered to account for the same amount of light. 
As such, the slight underestimation of the SMD by \citet{Driver18} with respect to our values may be due, at least in part, to the lower stellar masses derived by \textsc{MAGPHYS}. 
Our cSMD is compared against further SMD values from the literature in the bottom panel of Fig. \ref{fig:cSFRComparisonLiterature}, including the compilation of measurements by \citet{Madau14},\footnote{This compilation includes SMD measurements by \citet{Arnouts07, Gallazzi08, PerezGonzalez08, Kajisawa09, Li09, Marchesini09, Yabe09, Pozzetti10, Caputi11, Gonzalez11, Bielby12, Lee12, Reddy12, Ilbert13, Labbe13, Moustakas13, Muzzin13}} as well as \citet{Mortlock11, Santini12, Duncan14, Tomczak14, Grazian15, Song16, Davidzon17, Bhatawdekar19}. The agreement is generally close in all cases, suggesting that our SMD values are slightly overestimated in the first 2 billion years of the Universe.

\subsubsection{Comparison to Literature}

\begin{figure*}
	\centering
	\includegraphics[width=180mm]{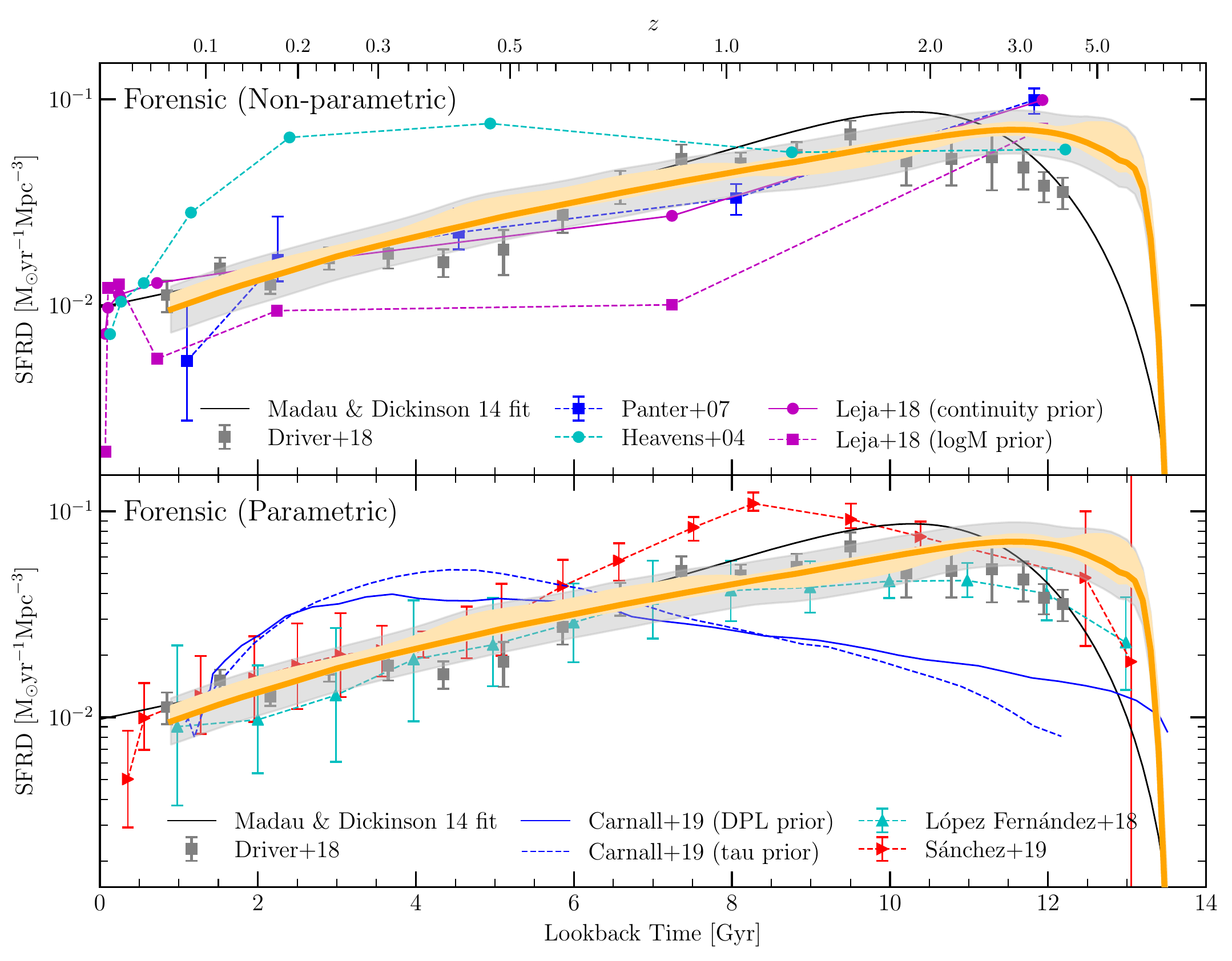}
	\caption{The cosmic star formation history from \textsc{ProSpect} (shown in orange), compared with other forensically determined star formation histories. The top panel compares our result directly with other non-parametrically determined CSFH curves \citep{Heavens04, Panter07, Leja18}, and the bottom panel compares our result with other parametrically determined CAFH curves \citep{Carnall19, LopezFernandez18, Sanchez19}. Note that we have converted the \citep{Sanchez19} SFRD from a \citet{Salpeter55} to a \citet{Chabrier03} IMF.  }
	\label{fig:cSFRComparisonForensic}
\end{figure*}

Numerous studies have previously attempted to recover the CSFH using forensic SED fitting. 
Of these studies, there is a mix of parametric and non-parametric approaches. 
The top panel of Fig. \ref{fig:cSFRComparisonForensic} compares our derived CSFH against previous results that used non-parametric SFHs \citep{Heavens04, Panter07, Leja18}, whereas the middle panel compares our CSFH against results from parametric SFHs \citep{Carnall19, LopezFernandez18, Sanchez19}. Note that from a visual inspection of the two panels, neither technique produces CSFH curves systematically more consistent with the observed CSFH. 

\citet{Heavens04} used \textsc{Moped} to recover the fossil record of $\sim$100,000 galaxies in the redshift range $0.0005 < z < 0.34$ from SDSS, resulting in a CSFH with a peak occuring much later than that measured by core-sample techniques. 
The SDSS data were later also analysed by \citet{Panter07}, who similarly used \textsc{Moped}, but recovered older stellar populations in their analysis that produce a CSFH more reflective of that observed via core-sampled methods. 
The non-parametric approach by \citet{Leja18} using the \textsc{Prospector} code, which also fitted SEDs of GAMA galaxies (noting that their CSFH has been scaled to the low-$z$ value), produces a very similar CSFH to that of \citet{Panter07}. Note that due to the low time resolution at high $z$ implemented by these non-parametric techniques, it would not have been possible to resolve the rising CSFH in the early Universe.  
\citeauthor{Leja18} emphasised that the greatest modelling aspect influencing the resulting star formation history is the selected prior. 
Fig. \ref{fig:cSFRComparisonForensic} highlights that regardless of the selected prior, however, the general trends that result are the same. 

While the parametric approach by \citet{Carnall19} does show the early-rising SFH, the peak occurs $\sim$ 6 Gyr later than that observed. 
The recent result by \citet{LopezFernandez18} is based on simultaneous SED- and spectral index-fitting for 366 galaxies in the CALIFA survey using \textsc{Starlight} with a parametric, delayed-$\tau$ parametrization of the SFH. 
Of the 9 parametrizations tested in their study, this was the model that was deemed to produce a CSFH most resembling that derived by observations. 
Despite the relatively small sample size used, this approach is able to recover the general shape of the CSFH, although we note that there was significant variation between the CSFHs derived by their separate parametrizations. 
Interestingly, the uncertainties are lower at higher lookback times than lower times in these results, which is counterintuitive given that older stellar populations are in general less well-constrained. 
Due to the relatively small galaxy sample, which results from a complex selection function and is not volume-limited, it is uncertain whether \citet{LopezFernandez18} use a cosmologically representative sample. As a result, the impact of the volume correction method on their derived CSFH is unclear. 
Finally, \citet{Sanchez19} built their CSFH using $\sim4,200$ galaxies from the MaNGA survey. A \citet{Salpeter55} IMF was used for this study, which we have corrected to a \citet{Chabrier03} IMF using a conversion factor of 0.63, as given by \citet{Madau14}. The general shape is consistent with the core-sample CSFH, albeit with a peak measured to be at $\sim8$ Gyr, slightly later than the observed peak by around 2 Gyr. 

The approaches used to model the metallicity of galaxies varied between these studies. For the \citet{Leja18} study using \textsc{Prospector}, both stellar- and gas-phase metallicity were fitted separately, using the stellar mass--metallicity relation as a prior. 
Similarly, \citet{Carnall19} used a constant-but-free metallicity in their study, utilising \textsc{Bagpipes} with a logarithmic prior between $0.2 < Z/\rm{Z_{\odot}} < 2$. Note that the lower limit of metallicity in \citet{Carnall19} is larger than our lower limit of $Z=10^{-4}$, corresponding to $0.05\,\rm{Z}_{\odot}$ -- a potential explanation for why the peak of the CSFH has been underestimated (as indicated in Fig. \ref{fig:MetallicityComparison}). 
\citet{LopezFernandez18} also fitted for a constant stellar metallicity, but because IFU data from CALIFA was used, this was done on a spaxel-by-spaxel basis, rather than at a global level. 
This extra constraint may be a contributing factor to the improved fit to the stellar populations of the last 8 Gyr, reducing the impact of a constant-metallicity assumption. 
We note that the recent 8 Gyr are well fitted by their analysis for most of the tested SFH parametrizations. 
\citet{Heavens04} recovered an average gas metallicity in each of the 11 time bins analysed in their work --- an approach very different to the other literature approaches outlined in this section. 

Due to the large galaxy sample analysed in our study, the uncertainty range presented for our CSFH measurement is smaller than that of \citet{LopezFernandez18} and \citet{Sanchez19}. Note, however, that this presented uncertainty only reflects the MCMC sampling error, and not other sources of uncertainty such as cosmic variance, SFH/ZH parametrization error, uncertainties in the stellar popularion models, et cetera.

\subsubsection{Stellar Mass Trends}
\label{sec:StellarMassTrends}

\begin{figure}
	\centering
	\includegraphics[width=85mm]{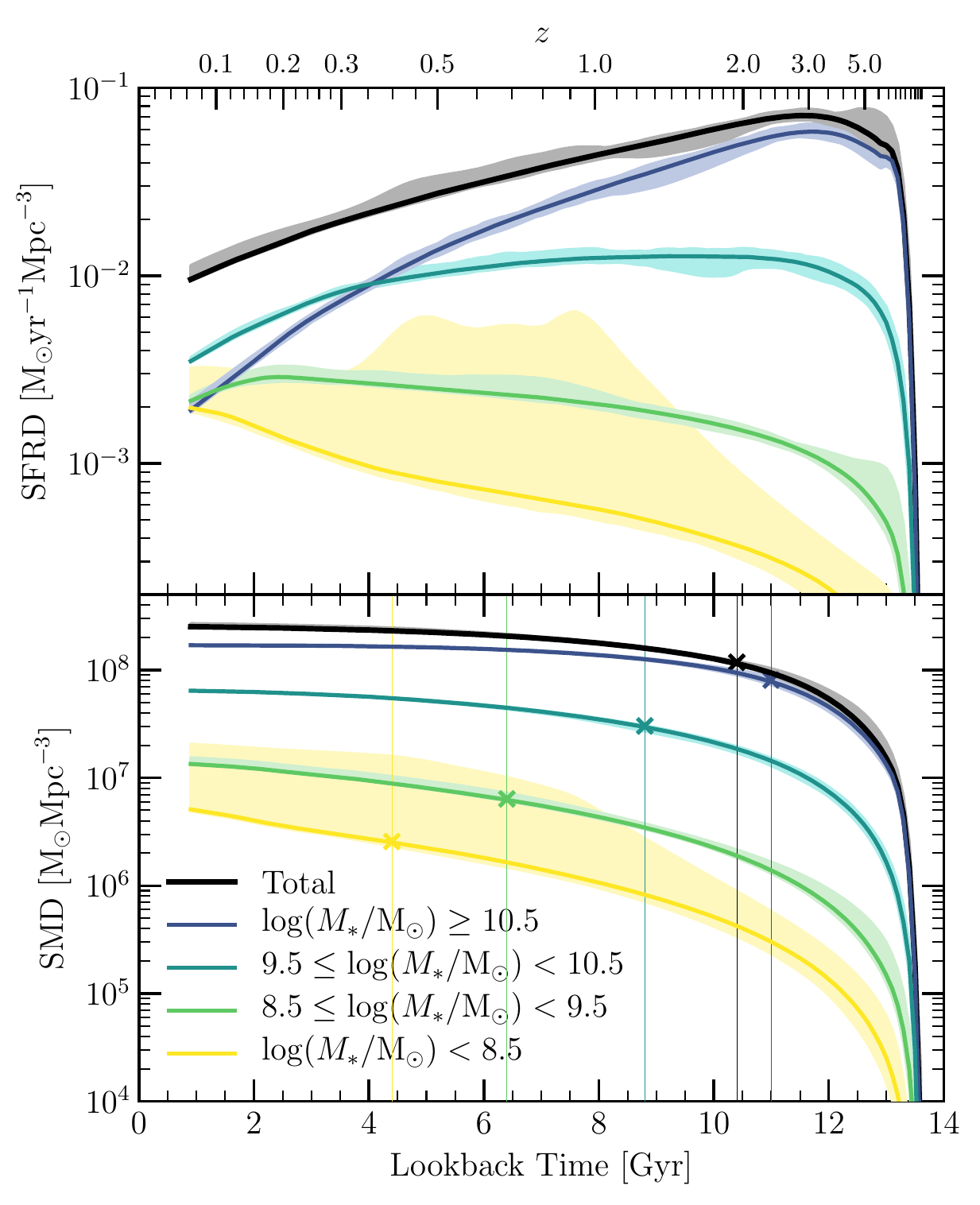}
	\caption{Top panel: Cosmic star formation history for the full sample in black, and the contributions of the individual present-day stellar mass bins in the coloured lines. Bottom panel: Cosmic stellar mass density, with the stellar mass bins divided into the same subcategories as in the top panel. For each subpopulation, we indicate the epoch at which 50 per cent of the stellar mass is formed with a cross and vertical line. }
	\label{fig:SFD-SMD_Mass}
\end{figure}

We now extract the contributions to the CSFH and SMD by galaxies in different stellar mass bins. Fig. \ref{fig:SFD-SMD_Mass} divides the CSFH and SMD into four stellar mass bins with boundaries at $\log(M_*/\rm{M}_{\odot})=$8.5, 9.5, 10.5.

While each stellar mass bin has a roughly equal contribution to the CSFH at local times, they differ in the early Universe, where the most massive galaxies today produced the most stars during cosmic noon. Correspondingly, the peak CSFH of each mass bin reduces with reducing stellar mass, as reflected by the decline of the total CSFH in the past 10 Gyr. This is entirely consistent with the ``downsizing" paradigm, which suggests that more massive galaxies formed their stars earlier \citep{Cowie96, Cimatti06, Thomas19}. This trend is qualitatively similar to that found by \citet{Heavens04} in their forensic analysis of SDSS galaxies. 
This downsizing is shown again in the bottom panel, where (as in Fig. \ref{fig:SFD-SMD_Morph}), the crosses and vertical lines indicate the epoch at which half of the stellar mass has formed. Note that for the most massive galaxies this time is earlier (11 Gyr ago), whereas for the least massive galaxies, this time is as recent as $\sim$4.4 Gyr ago. The values for the CSFH and SMD in Fig. \ref{fig:SFD-SMD_Mass} are tabulated in Table \ref{tab:SMDvalues}. 
These results are qualitatively consistent with a consensus developed by a variety of galaxy evolution studies over three decades or more, but it is remarkable that we are able to extract these global trends based only on our analysis of broadband SEDs for $\sim$7,000 galaxies at $z<0.06$. Note also that the priors implemented in our analysis are independent of the stellar mass of individual galaxies, and hence the fact that we recover the downsizing trend highlights that we are able to meaningfully extract qualitative differences in the SFHs of different galaxy samples.

The contributions of these bins to the SMD scales directly with the stellar masses. 
Note that our sample is only mass complete down to $\log(M_*/\rm{M}_{\odot}) = 9$, and hence the lowest mass bin is more prone to bias due to our adopted mass correction.

We compare our results to those implied by the cosmological models \textsc{Shark}, \textsc{EAGLE}, and IllustrisTNG in Appendix \ref{sec:ModelComparison}. 

\subsubsection{Morphological Trends}

\begin{figure}
	\centering
	\includegraphics[width=85mm]{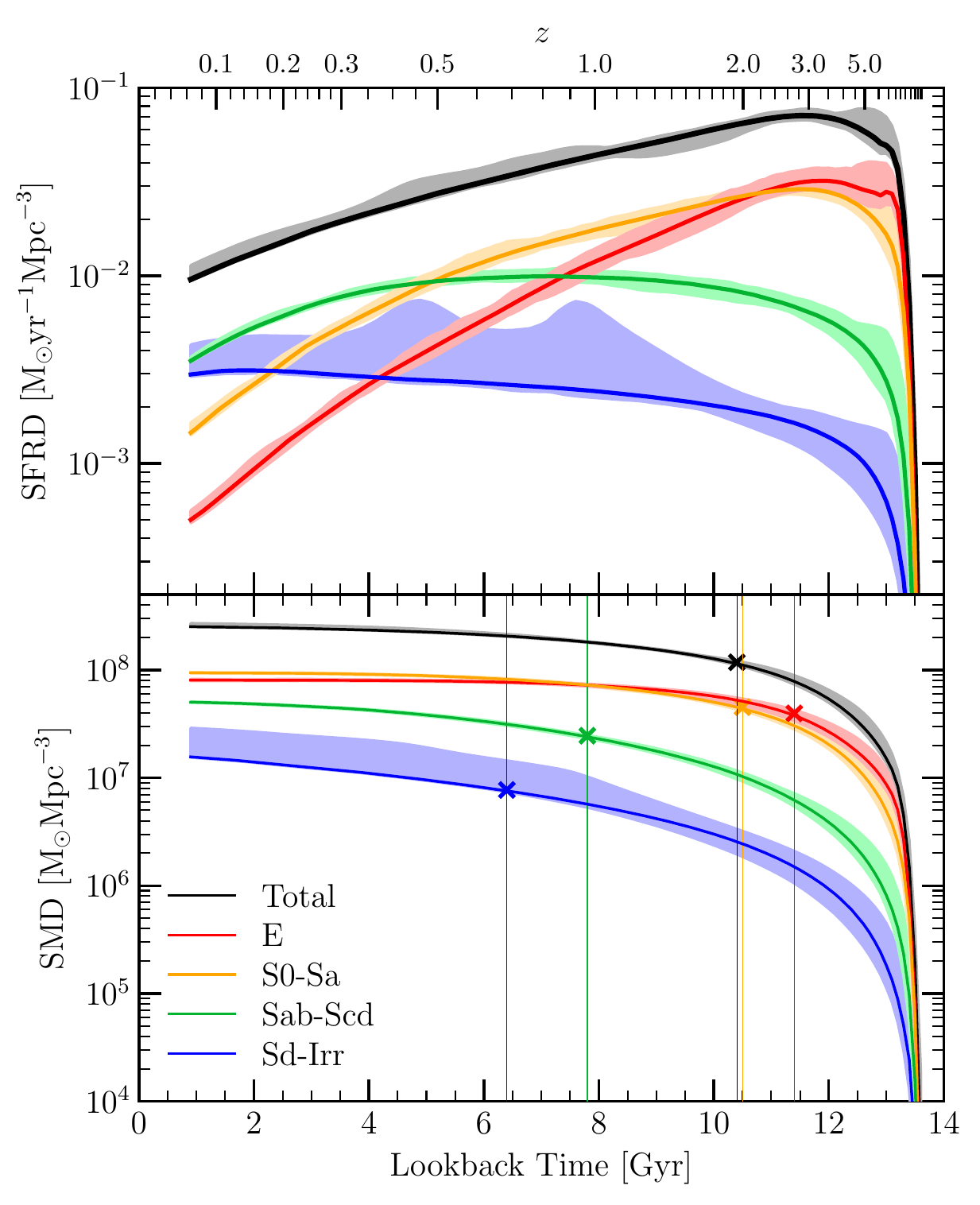}
	\caption{Top panel: Cosmic star formation history for the full sample in black, and the contributions of the individual galaxy morphologies in the coloured lines. Bottom panel: Cosmic stellar mass density, with the morphologies divided into the same subcategories as in the top panel. For each subpopulation, we indicate the epoch at which 50 per cent of the stellar mass is formed with a cross and vertical line.  }
	\label{fig:SFD-SMD_Morph}
\end{figure}

In a similar fashion to the previous section, we are able to utilise the morphological classifications of individual galaxies in our sample to recover the corresponding CSFH for broad morphological classes, taking our resulting SFHs at face value. This allows us to study the build-up of mass that results in the different morphologies (according to their visually classified Hubble types) we observe in the Universe today. 

The populations of galaxies with specific morphological classifications at high redshifts, however, are not necessarily the direct progenitors of populations of galaxies with the same morphological classifications at low redshifts, as morphologies of individual galaxies are expected to change with time \citep[as noted in many studies, including for example][ Hashemizadeh et al. in prep]{Dressler80, Lackner13, Bizzocchi14}. Consequently, it is very difficult to directly probe the evolution of populations of galaxies with a core-sample observational approach, and hence our forensic approach is well suited for studying mass growth as a function of present-day morphology. 

The contributions towards the CSFH by each of the four main morphological types (\texttt{E}, \texttt{S0-Sa}, \texttt{Sab-Scd} and \texttt{Sd-Irr}) are indicated in Fig. \ref{fig:SFD-SMD_Morph} by the red, orange, green and blue solid lines respectively.
At the earliest times in the Universe, star formation predominantly occurs in present-day \texttt{E} and \texttt{S0-Sa} (mostly early-type, but also including \texttt{Sa} galaxies) galaxies, whereas at recent times star formation dominates in \texttt{Sab-Scd} and \texttt{Sd-Irr} (late-type) galaxies. This supports our understanding that early-type galaxies are generally old, whereas late-type galaxies are younger. 
In their forensic analysis of the histories of CALIFA galaxies, \citet{LopezFernandez18} examined the evolution of the SFHs of galaxies with different morphoplogies in their sample, and did not identify a clear difference in the epoch of peak star formation of galaxy populations with different morphologies. Rather, the SFHs of different morphologies were separated mainly by the amount of star formation present within each population, with early-type galaxies having the highest star formation rates, and late-type galaxies having the lowest star formation rates up until recently. Only in the last $\sim$2 Gyr were \texttt{Sab-Scd} and \texttt{Sd-Irr} galaxies observed to have higher SFRs than \texttt{E} and \texttt{S0-Sa} galaxies. 
\citet{Guglielmo15} measured the SFHs of galaxies in both the field and cluster environment by applying a spectrophotometric code adapted from \citet{Poggianti01} to the PM2GC and WINGS datasets to derive the contributions of late-type, lenticular and elliptical galaxies to the CSFH in each environment. They show that, in the field environment, the CSFH is dominated by late-type galaxies at all epochs. In the cluster environment, however, present-day early-type galaxies dominate the CSFH at $z>0.1$, whereas present-day late-type galaxies only dominate in the most recent time bin. These trends observed in the cluster environment are qualitatively consistent with the overall trends by morphology observed in Fig. \ref{fig:SFD-SMD_Morph}. 

The contributions from different morphological types to the cosmic stellar mass density are shown in the bottom panel of Fig. \ref{fig:SFD-SMD_Morph}.
This figure would indicate that the progenitors of current-day \texttt{S0-Sa} galaxies have dominated the mass budget of galaxies since a lookback time of $\sim$ 10 Gyr. As is shown in Fig. \ref{fig:StellarMassHist}, the reason for this is that the individual galaxies of this morphological class have higher masses on average, and not because there is a surplus of \texttt{S0-Sa} galaxies. 
In the earliest epoch of the Universe (earlier than 10 Gyr ago), the contributions to the SMD by \texttt{E} and \texttt{S0-Sa} galaxies are consistent within sampling uncertainty. 
We have indicated the epoch at which 50 per cent of the stellar mass is formed with a cross and vertical line in this panel. For the full sample, 50 per cent of stars have already formed by just over 10 Gyr ago. For each morpholgical subpopulation, this point varies, from $\sim$11.4 Gyr for \texttt{E} galaxies, to $\sim$6.4 Gyr for the \texttt{Sd-Irr} sample.  The values for the CSFH and SMD in Fig. \ref{fig:SFD-SMD_Morph} are tabulated in Table \ref{tab:SFRDvalues}. 

Despite the difficulties caused by progenitor bias in comparing our results to those of high-redshift studies, which measure the properties of galaxies with different morphologies over a range of epoch, we take this opportunity to reflect on observational findings in the literature, and how they compare with our results. 
The relative contributions of galaxies with bulges (an alternative way of viewing morphoplogy) was studied by \citet{Grossi18} at $z<1$ for galaxies in the COSMOS survey. This study showed that in this redshift range the CSFH is lower (and drops off faster) for galaxies that are ``bulgy" than for galaxies that are ``bulgeless".  Likening ``bulgy" galaxies with early-types, and ``bulgeless" galaxies as late-types, this result is qualitatively consistent with that seen in the top panel of Fig. \ref{fig:SFD-SMD_Morph}. 

\citet{Tamburri14} analysed the different contributions of massive, $\log(M_*/\rm{M}_{\odot}) > 11$, early- and late-type galaxies to the SMD, finding that the contribution of late-type galaxies is greater than that of early-type galaxies at all times in $0.5 < z < 2$ (their fig. 13), and that the evolution in the SMD of early-type galaxies was much stronger than that of the late-type galaxies. 
The SMD was also the target of a study by \citet{Ilbert10}, who divided the SMD by the contributions of galaxies with varying levels of star formation activity, showing that the assembly of massive early-type galaxies has not occurred until $z\sim1$. 

Note that the two studies by \citet{Tamburri14} and  \citet{Ilbert10} both suggest that the SMD is dominated by late-type galaxies at all epochs studied, which is exactly the oppostive of what we see in the lower panel of Fig. \ref{fig:SFD-SMD_Morph}, where the early-type morphoplogies clearly dominate the SMD at all points in history. It is critical to note that this comparison highlights the consequence of progenitor bias --- and the reason why observations of galaxies with morphologies at high $z$ cannot be directly compared with a forensic-type analysis based on $z=0$ morphologies. Those star-forming galaxies that were observed to dominate the SMD at high $z$ are likely to have quenched in the following time, and would have been included in our sample as early-type galaxies. 

A powerful tool that can be used to circumvent progenitor bias in overcoming the differences between ``direct" and ``forensic" observational techniques is that of simulations. 
\citet{Martin18} used the Horizon-AGN simulations to try to deal with progenitor bias by identifying the progenitors of modern-day early-type galaxies in the simulations. They find that by $z\sim1$ around 60 per cent of the mass within current massive early-type galaxies had already been formed. 
Additionally, they note that the effect of progenitor bias is significant --- beyond $z\sim0.6$ less than half of early-type galaxy progenitors have early-type morphologies, highlighting why observational studies like those of \citet{Tamburri14} and \citet{Ilbert10} measure a SMD dominated by late-type galaxies at higher redshifts, while we note that the SMD is dominated by the progenitors of modern-day early-type and Sa galaxies. 

\citet{Trayford19} used the EAGLE simulations to derive results in a similar way to this work, by dividing the SFRD into contributions by disc, spheroid and asymmetric galaxies with $M_*>10^{9}\rm{M}_{\odot}$. In their study, the peak CSFH contribution for disc galaxies was shown to be at $\sim$ 8 Gyr, which is much earlier than the broad peak we find for spiral galaxies between $\sim 4-10$ Gyr ago, however the agreement would be much better if we also include lenticular galaxies, which also contain a disc, in general. The broader peak that we identify may also be due to the contribution of lower-mass galaxies, which typically have SFHs that peak at more recent times (see the discussion in the next section). We note that a direct comparison here is difficult due to the different definitions of morphology. 

We leave an analysis of the individual contributions of bulges and discs separately to the CSFH and SMD analogously to \citet{Driver13} for future work. 

\begin{figure}
	\centering
	\includegraphics[width=85mm]{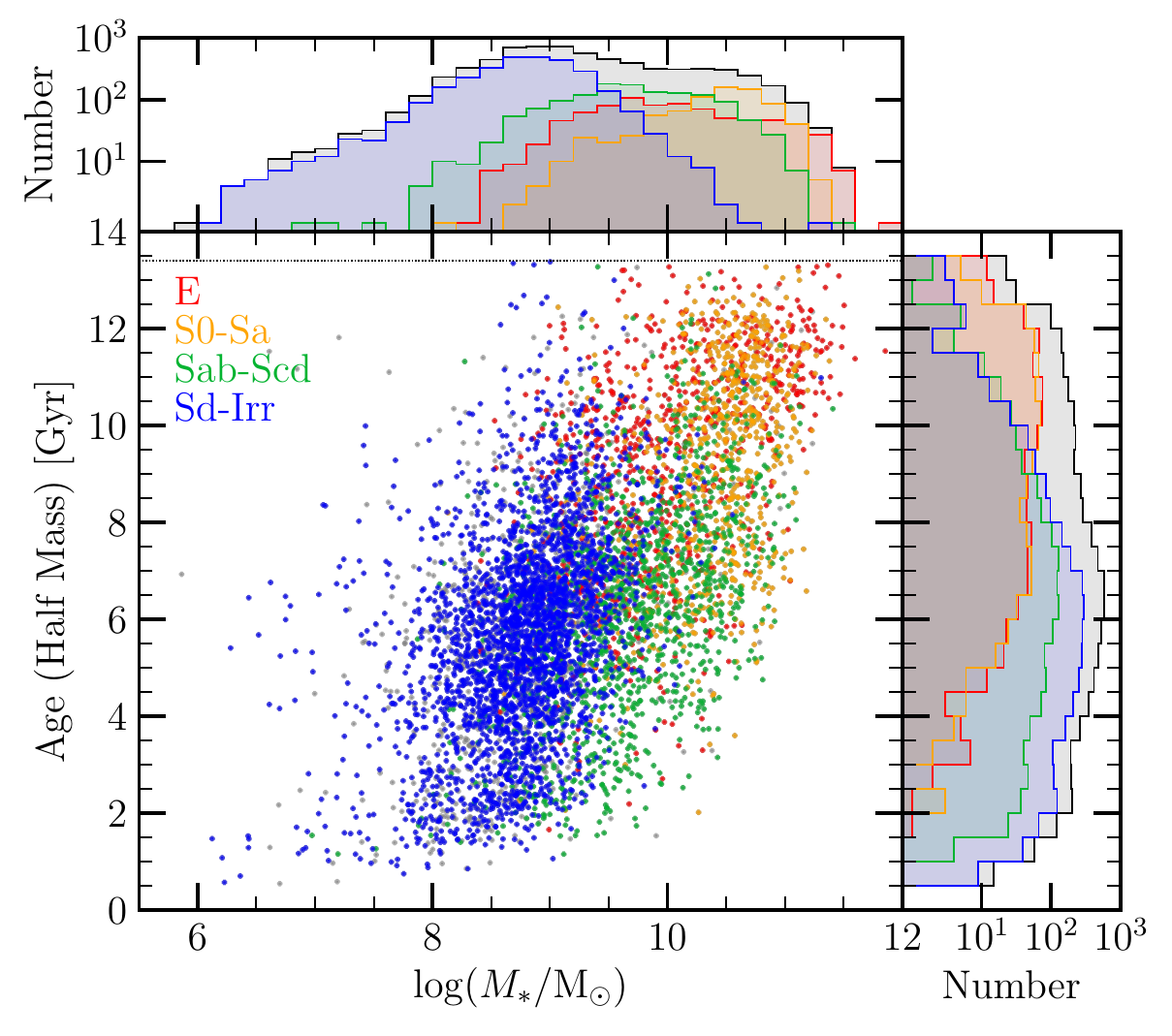}
	\caption{Distribution of the age at which 50 per cent of the stellar mass of individual galaxies is formed, against stellar mass, coloured by the morphology of the objects. The top histogram shows the distribution of stellar masses, separated by morphology, and the right histogram shows the distribution of the half mass age values, also separated by morphology. The black dotted line shows the maximum age of the Universe. }
	\label{fig:MassPeak}
\end{figure}

The effects of stellar mass and morphology cannot be fully disentangled. We show in Fig. \ref{fig:MassPeak} how the age at which half the stellar mass is formed varies with stellar mass. The downsizing of the sample is clearly seen here. Note that there are trends with morphology here as well, as earlier-type morphologies tend also to be more massive.

\section{Metallicity}
\label{sec:Metallicity}

Due to the closed-box implementation of the gas-phase metallicity in this work, our resulting data set also allows us to analyse the evolution in gas-phase metallicity across the sample over cosmic time. 

\begin{figure}
	\centering
	\includegraphics[width=85mm]{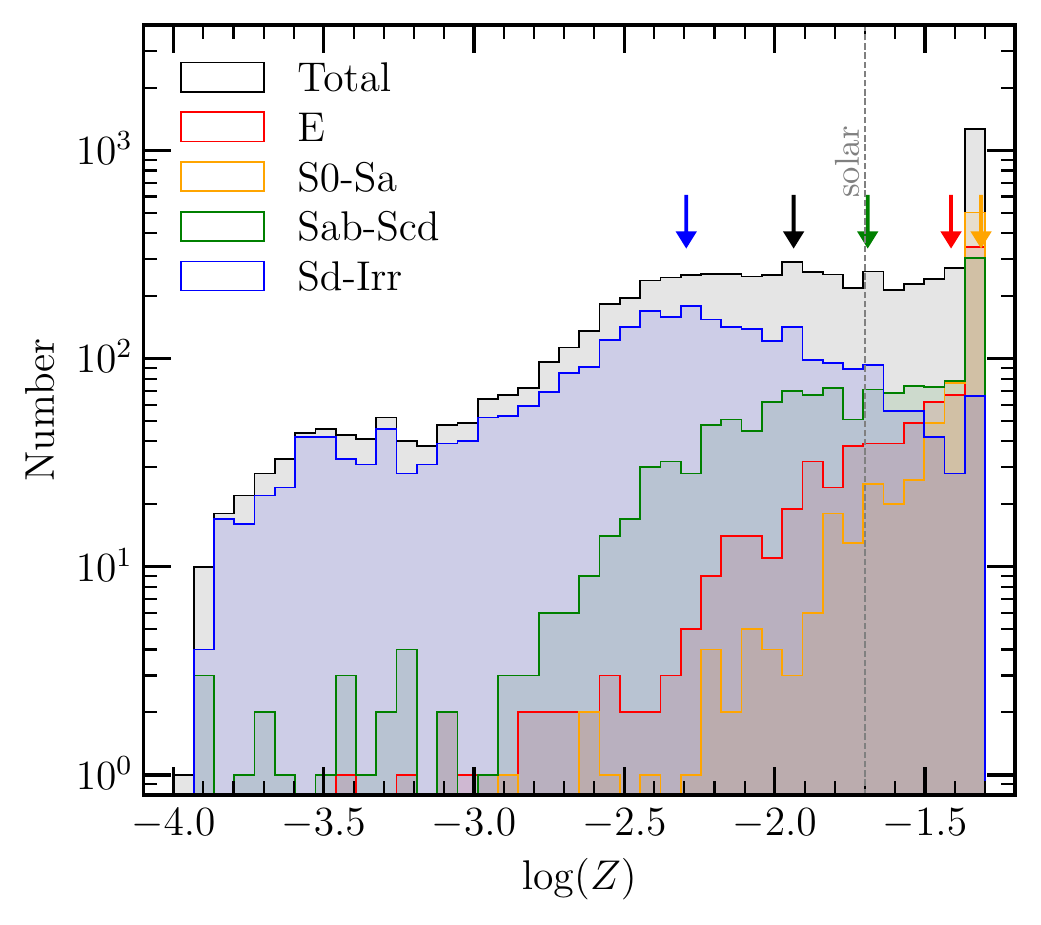}
	\caption{Distribution of gas-phase metallicities derived for GAMA galaxies, separated by morphology. For each morphological group, the median metallicity is shown with an arrow. Note that the median metallicity for the \texttt{S0-Sa} galaxy population is limited by the limit of the \citet{Bruzual03} templates. }
	\label{fig:MetallicityDistribution}
\end{figure}

We show the distribution of derived gas-phase metallicities of the sample as divided by morphology in Fig. \ref{fig:MetallicityDistribution}. 
This figure shows that early-type galaxies tend to have higher present-day metallicities, whereas late-type galaxies have lower metallicities, as one would expect. 
For each population, the median metallicity is indicated by an arrow. 
Note that \texttt{S0-Sa} galaxies have a median metallicity consistent with the upper limit of the \citet{Bruzual03} templates.

\subsection{Metallicity evolution}

\begin{figure}
	\centering
	\includegraphics[width=85mm]{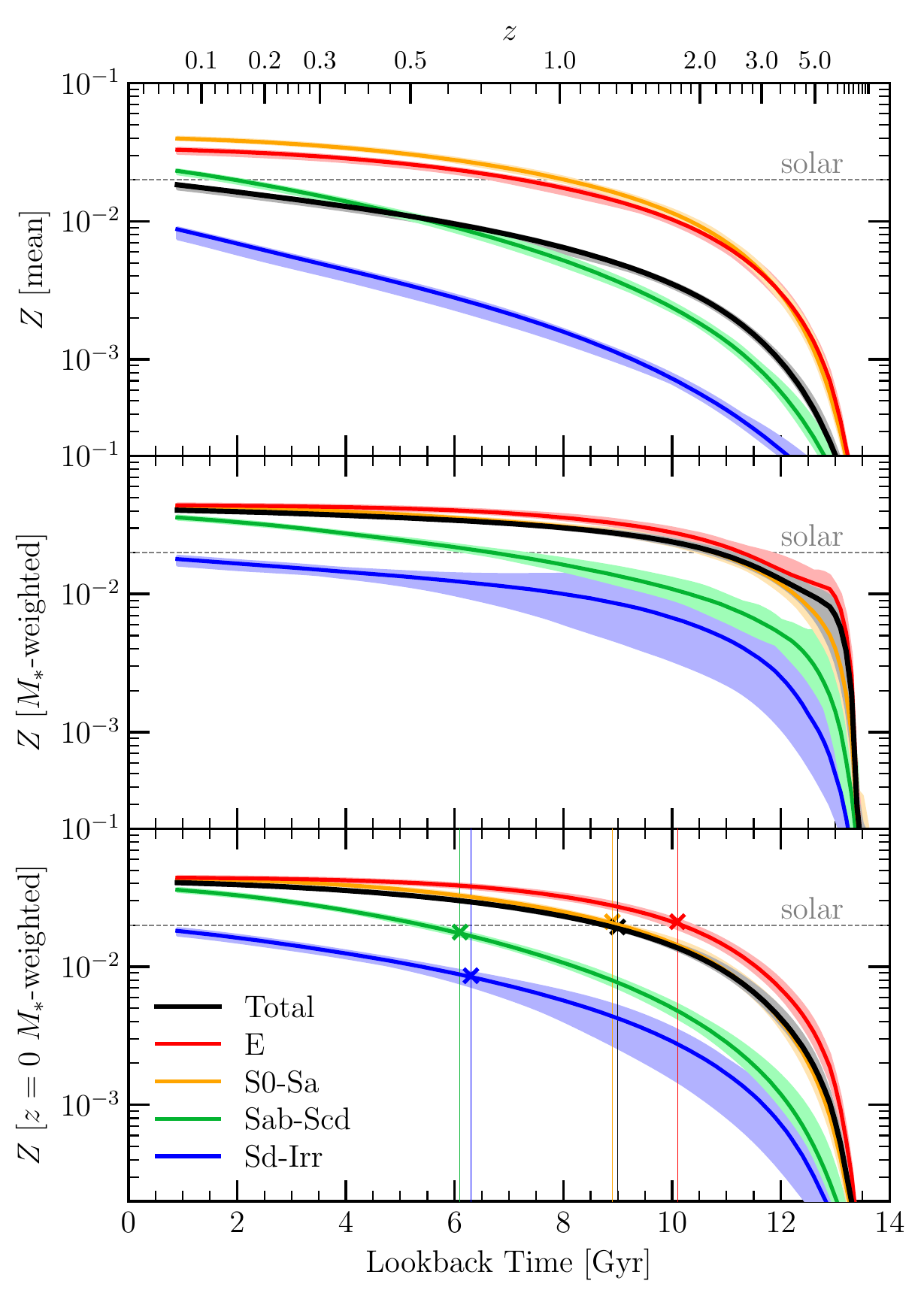}
	\caption{Top panel: mean gas-phase metallicity evolution for the full sample of galaxies (black), and also divided by morphological classification. Middle panel: Local stellar mass-weighted metallicity evolution, for the same populations as in the top panel. Bottom panel: Present-day stellar mass-weighted metallicity evolution, for the same populations as in the top panel. In each panel, solar metallicity is shown with a horizontal grey line. }
	\label{fig:MetallicityEvolutionMorph}
\end{figure}

Fig. \ref{fig:MetallicityEvolutionMorph} follows the evolution of metallicity in our sample, subdivided by visual morphological classification, calculated in three different ways. 
As with the other plots presented in this paper, trends for the full sample are shown in black, \texttt{E} in red, \texttt{S0-Sa} in orange, \texttt{Sab-Scd} in green, and \texttt{Sd-Irr} in blue. 
The top panel shows the mean metallicity evolution for each population. 
Due to the mass-dependent nature of metallicity, this representation of the evolution is influenced by the mass completeness of the samples. 
The middle and bottom panels, conversely, show the stellar mass-weighted evolution of the metallicity, indicative of the mean metallicity per unit of stellar mass. 
The middle panel shows the resulting metallicities that are weighted by the corresponding stellar mass of the galaxies at that epoch. This is a reflection of the typical mean metallicity per unit mass throughout cosmic time.
The bottom panel shows the metallicity evolution weighted by the $z=0$ stellar mass of each galaxy. This representation is a clearer depiction of the typical metallicity histories of individual populations of galaxies. 

Each panel shows that the build-up of metallicity is rapid in the early Universe, where star formation is most prolific, and comparably slow in recent times.  Note that, as a result of our closed-box prescription, the shape of the mean metallicity evolution is very similar to that of the SMD in the lower panel of Fig. \ref{fig:SFD-SMD_Morph}.
While on average, our results suggest that \texttt{S0-Sa} galaxies are more metal-rich than \texttt{E} galaxies (as shown both in Fig. \ref{fig:MetallicityDistribution} and by the top panel of Fig. \ref{fig:MetallicityEvolutionMorph}), per unit of star formation, \texttt{E} galaxies are shown to be more metal-rich than \texttt{S0-Sa} galaxies in the middle and bottom panels of Fig. \ref{fig:MetallicityEvolutionMorph}. This is due to the difference in stellar mass distributions of \texttt{E} and \texttt{S0-Sa} galaxies, where the most massive \texttt{E} galaxies are more massive than the most massive \texttt{S0-Sa} galaxies. It should be noted however that the differences between the trends for these populations are small, and given the limitations of the SFH and metallicity history parametrizations, these differences are probably not significant. 

We reiterate at this point that these metallicity evolution profiles have been derived using a closed-box metallicity model, neglecting any potential impact from gas inflows or outflows. Phenomena such as these will inevitably impact the evolving metallicity of galaxies, as demonstrated by \citet{Edmunds97}.

\subsection{Cosmic metal density}

Through implementation of a closed-box metallicity evolution model, we are hence able to infer the corresponding metal mass present in both the gas and the stars separately at each epoch for each individual galaxy, as given by our parametrization. The metal mass in gas is calculated as:
\begin{equation*}
M_{Z, \rm{gas}}(t) = \it{Z}(t)\times \it{M}_{*, \rm{total}}(t)\times \it{f}_{\rm{gas}}(t), 
\end{equation*}
where the gas fraction, $f_{\rm{gas}}(t)$, prescribed by \citet{Robotham20}:
\begin{equation*}
f_{\rm{gas}}(t)=e^{-(\it{Z}(t)-\it{Z}_{\rm{initial}})/\rm{yield}}. 
\end{equation*}
Here, $Z_{\rm{initial}}$ represents the starting metallicity (set to $10^{-4}$), and $M_{*, \rm{total}}$ represents the cumulative stellar mass formed by that time. We highlight that $M_{*, \rm{total}}(t)$ is greater than the actual stellar mass of the galaxy at that time, which we denote $M_{*, \rm{remaining}}(t)$ to represent the stellar mass of the galaxy remaining after mass loss. 

The metal mass in stars is calculated as:
\begin{equation*}
M_{Z, \rm{stars}}(t) = \frac{\sum\limits_{\rm{age}=0}^t \it{Z}(\rm{age})\rm{SFR}(\rm{age})}{\sum\limits_{\rm{age}=0}^t \rm{SFR}(\rm{age})}  \times \it{M}_{*, \rm{remaining}}(t), 
\end{equation*}
where $M_{*, \rm{remaining}}(t)$ is the stellar mass at each epoch. 

\begin{figure}
	\centering
	\includegraphics[width=85mm]{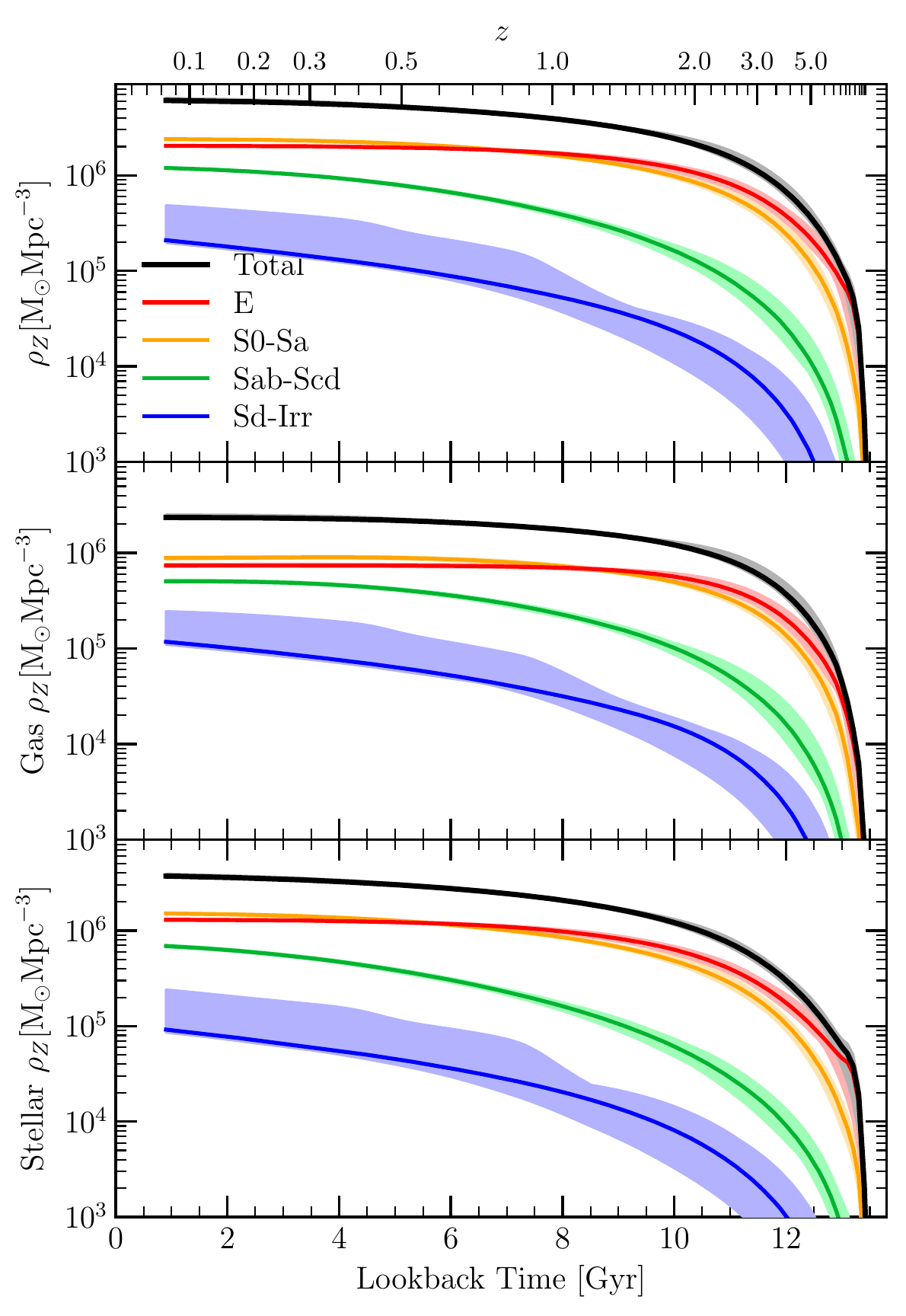}
	\caption{The derived total cosmic metal mass density (top panel), metal mass density in gas (middle panel), and metal mass density in stars (bottom panel). }
	\label{fig:CosmicMetalDensity}
\end{figure}

Applying the same technique to stack, mass-correct and volume-correct these profiles as done to generate the CSFH and SMD, we build the cosmic metal mass density for the total metals, as well as the gas metals and star metals separately, in Fig. \ref{fig:CosmicMetalDensity}.  
The top panel of the plot shows the total metal mass density (including both stars and gas), whereas the gas and stellar metal components are broken down in the middle and bottom panels respectively. 

\begin{figure*}
	\centering
	\includegraphics[width=180mm]{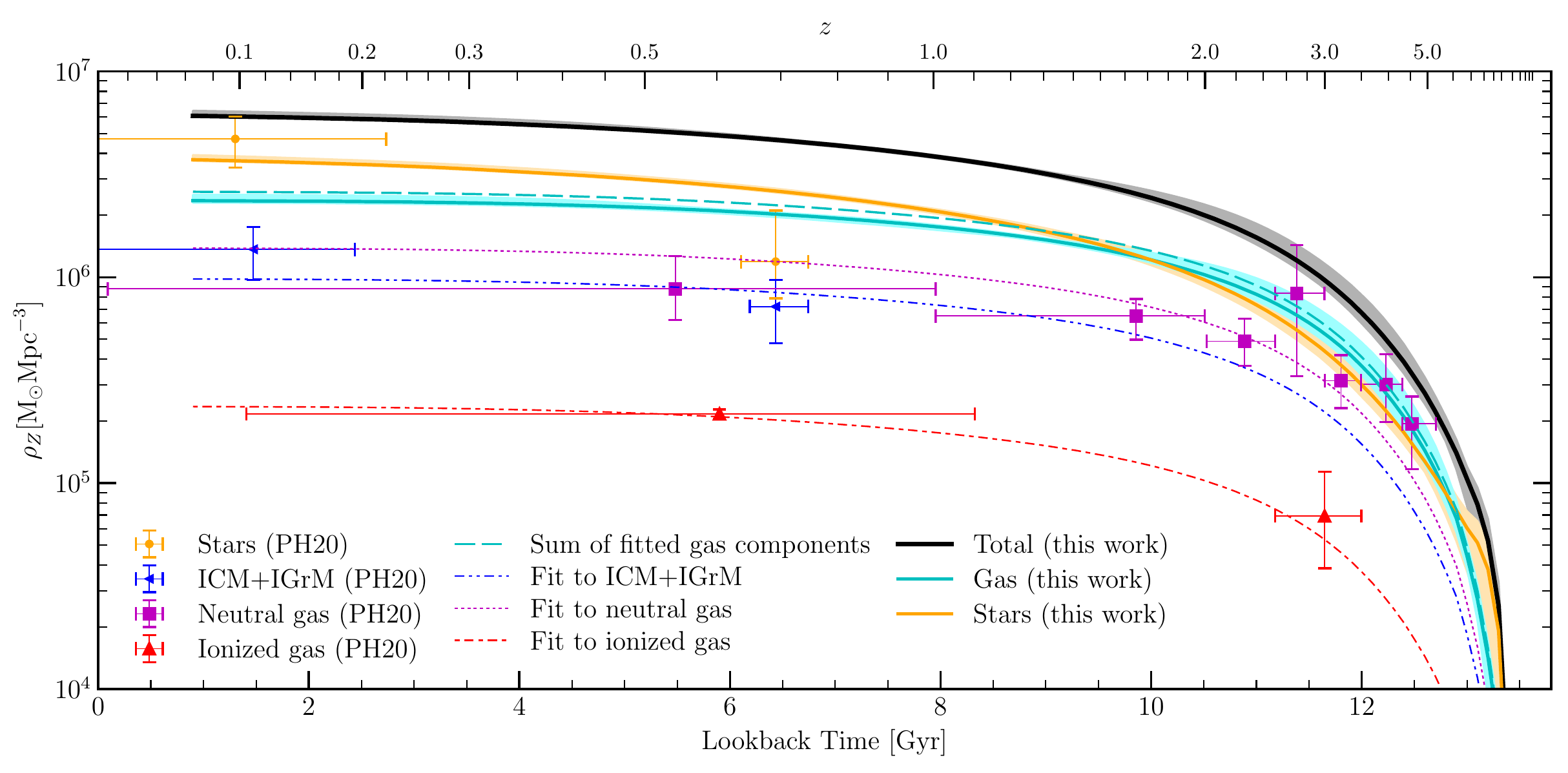}
	\caption{The derived total cosmic metal mass density, including the contribution by gas and stars within galaxies, as compared with data from \citet{Peroux20}. 
		In order to facilitate a comparison between the gas metal mass density derived in this work and observational values, we scale the total metal mass density derived in this work to match the observational points for each gas component, as a rough fit to the data. The sum of these fits is shown in the dashed cyan line. We note that this is broadly consistent with the solid cyan line.  }
	\label{fig:CosmicMetalDensityLiterature}
\end{figure*}

We compare the metal density outputs from our analysis with those presented in the literature in Fig. 
\ref{fig:CosmicMetalDensityLiterature}. Here, the total cosmic metallicity density is shown in black, whereas the stellar component is shown in orange, and the gas component in cyan. 
The values for these curves are tabulated in Table \ref{tab:MetalDensityvalues}. 
We show observational measurements from \citet{Peroux20} for stars, ionized gas, intracluster/intragroup medium (ICM+IGrM), and neutral gas  in Fig. \ref{fig:CosmicMetalDensityLiterature}. 
We note that the metal mass density of the gas we derive is not directly comparable to any one of the observed gas components, but rather it likely represents a sum of all components. 
In order to judge whether the gas metal mass density we derive agrees with the sum of the observed components, we roughly fit the trends of each observed component, shown as dashed lines. This fit is conducted by scaling the shape of the total metal mass density, shown in black in the plot, until it passes through the observed data points. The sum of these fits is shown in the plot as a dashed cyan line. This line agrees relatively well with the gas metal mass density we derive, shown as the solid cyan line. 
While this approach is very rough, and grants some degree of agreement by construction, it shows that the metal mass we derive is as consistent with observations as one could hope from the style of analysis employed in this paper. 
Because we assume a closed-box metallicity evolution, the gas metal mass in our study represents all the metals formed within the galaxy that do not remain locked in stars. 

\section{Caveats}
\label{sec:Discussion}

As with any implementation of SED fitting, there are numerous assumptions that have been made in this analysis, and hence we take this opportunity to address a number of caveats in this section. 

The critical assumption in modelling the SFHs of individual galaxies via their SEDs is that any galaxy has only had a single progenitor at any point in history. 
In reality, mergers between galaxies are known to occur (particularly for massive galaxies, where over half of the mass could be formed ex-situ, rather in-situ, as shown by \citealt{Bellstedt18, Forbes18}), and hence this assumption cannot be true for the whole sample. 
While the impact of this will be greatest for the construction of stellar mass functions at earlier epochs (which we have not quantified here) it is also possible that this assumption severely limits our ability to constrain SFHs for some individual galaxies. 
If the SFHs for two galaxies that are destined to merge are vastly different, then the SFH of the ``descendent" galaxy cannot be adequately described by a unimodal SFH, as we have done in this analysis. 
As a result, we expect that a unimodal SFH would be badly constrained for such a galaxy. 

Another consequence of galaxy mergers is that the assumption of a closed-box metallicity evolution becomes even less appropriate, as it is possible for either metal-rich or metal-poor gas to enter the system through a merger or interaction between galaxies (as observed by, for example, \citealt{Pearson18, Serra19} and shown in simulations by \citealt{Iono04}). 
Furthermore, the stars formed in the progenitor galaxies formed in their own gas reservoirs, which can have different metallicities. In addition, the assumption of closed-box metallicity evolution does not allow for gas inflows or outflows, which could also alter the gas-phase metallicity. This was highlighted by \citet{LopezSanchez10}, who showed that measured oxygen abundances were not consistent with the predictions from a closed-box chemical enrichment model for a sample of 20 starburst galaxies. 
While the implementation of the metallicity evolution in this paper is a significant improvement over the typical approach in the literature, it is not an exhaustive solution. 
In a future paper, we will explore the mass--metallicity relation and its evolution with cosmic time, resulting from this \textsc{ProSpect} analysis. 

We believe that the impact of the above two assumptions, while potentially large on individual galaxies, is minimal when assessing trends in galaxy evolution at a statistical level. This is supported by a test conducted by \citet{Robotham20}, in which the SEDs generated by the semi-analytic model \textsc{Shark} \citep{Lagos19} were fitted by \textsc{ProSpect}, to recover the SFH. While the burstiness of the true SFH could not be recovered by the parametric SFH, it was found that the stacked SFH of the galaxy population could be well recovered (see fig. 29 of \citealt{Robotham20} for an illustration). 

In this first implement of \textsc{ProSpect} modelling, we do not include nebular emission lines in our modelling. We expect that this does not have an impact on our results, due to the broadband nature of our photometry. In order to quantify this, we use the derived CSFH from this work as an input to \textsc{ProSpect}, to model the total sample SED with and without emission line features \citep[see][for a description of this feature in \textsc{ProSpect}]{Robotham20}. We find that the fluxes in the photometric bands used in this work increase on average 0.4 per cent, up to a maximum of 2 per cent, when including emission lines. In all cases the implemented error floor (shown in Table \ref{tab:PhotometrySummary}) is significantly larger than the relative increase caused by emission lines, highlighting that any impact from emission lines on our photometry will have been absorbed by the applied error floor. For data analysis using narrow photometric bands at higher redshift, the effect of emission lines may need further consideration.   

We have also not accounted for the potential presence of AGN at all in our approach, which would impact on our ability to accurately model the SED, if an AGN is present. We expect the number of AGN in our sample to be very low, \citep[likely fewer than 30 galaxies, according to a study of AGN in GAMA by][]{Prescott16}, and we expect that such a small number will not have a significant impact on our final CSFH and SMD results. For those individual galaxies with significant AGN emission, the determined properties and SFHs are potentially biased. We note, however, that the bulk of AGN emission occurs in the mid-IR portion of the SED, where photometric uncertainties and modelling floors provide little constraining power to the final SED fit. As such, AGN emission will result in larger mid-IR residuals observed in our fitting, without having a large impact on the derived star formation properties of the galaxy. The minimal AGN presence in our sample is highlighted by the fact that a significant population of galaxies for which the mid-IR flux is underestimated by \textsc{ProSpect} is not identified in this work. On the contrary, the \textsc{ProSpect} modelled fluxes in the mid-IR tend to be, on average, slightly in excess of the observed fluxes, hence we do not expect any serious impact of un-modelled AGN emission in this regime.

\section{Summary}
\label{sec:Summary}

We have applied the SED-fitting code \textsc{ProSpect} to extract the star formation histories of 6,688 galaxies at $z<0.06$ in the equatorial fields of the GAMA survey. Through the implementation of a parametric SFH, and a closed-box implementation of the metallicity evolution that takes its shape from the derived SFH, we have shown that we are able to loosely recover the cosmic SFRD measured by direct observations of galaxy SFRs over ranging redshifts (shown in Fig \ref{fig:cSFRComparisonLiterature}). 
We stress that a physically motivated implementation of the metallicity evolution is essential in order to extract the SFRD with the correct shape and position of the peak, and therefore correctly estimate the SFHs of individual galaxies. 

We have been able to assess the differential contribution to the SFRD and SMD by galaxies with different present-day stellar masses (Fig. \ref{fig:SFD-SMD_Mass}). As is expected, we find that the most massive galaxies peaked in their SFRD earlier, and had a larger contirbution to both the SFRD and (somewhat by definition), the SMD. These results directly support the ``downsizing" paradigm. 

Similarly, by using visual classification of the Hubble type morphologies of these galaxies, we are able to extract the contribution to the SFRD and SMD by galaxies with different present-day morphologies (Fig. \ref{fig:SFD-SMD_Morph}). This analysis shows us that the SFRD in the very early Universe was dominated by present-day elliptical galaxies, but by a lookback time of 10 Gyr, the progenitors of present-day \texttt{S0-Sa} population made the largest contribution to the SFRD. These two populations combined, are largely responsible for the peak in the SFRD at so-called ``cosmic noon" at $z\sim2$. 
The majority of mass formed in the first $\sim4$ Gyr of the Universe (85 per cent) ultimately ends up in \texttt{S0-Sa} or elliptical galaxies. 
Late-type galaxies became dominant in the SFRD much later in the Universe, with present-day spiral galaxies experiencing a very broad peak in their contribution to the SFRD between lookback times of 4--10 Gyr, with a contribution dominating over \texttt{S0-Sa} galaxies for the first time at a lookback time of $\sim$ 4 Gyr. Present-day \texttt{Sd-Irr} galaxies have slowly increased their contribution to the SFRD with time, experiencing their peak at the present day. These results support the general consensus that early-type galaxies formed earlier in the Universe and are dominated by older stellar populations, whereas late-type galaxies have built up their stellar mass more recently, and as such have younger stellar populations in general. 

Due to the selected metallicity evolution implementation, we are able to extract the metallicity evolution for the full sample, as well as for each subpopulation by visual morphology (Fig. \ref{fig:MetallicityEvolutionMorph}). This suggests that most of the build-up of metals occurs in the early Universe, with early-type galaxies forming their metals faster than late-type galaxies (reaching half their final gas-phase metallicities by $\sim$ 9 Gyr ago for early-types versus $\sim$ 6 Gyr ago for late-types). Additionally, we find that the present-day gas-phase metallicities are greater in elliptical and \texttt{S0-Sa} galaxies (around twice solar, see Fig. \ref{fig:MetallicityDistribution}), whilst they are lower for \texttt{Sab-Scd} (just above solar) and \texttt{Sd-Irr} (around half solar) galaxies. 

Finally, our closed-box metallicity implementation allows us to convert our metallicity evolution measurements into a cosmic metal mass density evolution for both the stellar and gas components separately (Fig. \ref{fig:CosmicMetalDensity}).
Additionally, we present the relative contribution of stars and gas to the total metal content in the Universe, showing that metals in stars only dominate after a lookback time of $\sim$9 Gyr (Fig. \ref{fig:CosmicMetalDensityLiterature}). A preliminary comparison to observational data shows that the total metal mass in gas from this work is broadly consistent with the sum of metal mass found in ionized gas, neutral gas and intracluster/intragoup medium. This suggests that our method produces a reasonable quantity of metals, although the closed-box metallicity approach (neglecting phenomena such as gas inflows and outflows), provides no information as to the location of these metals. 
A detailed analysis of the mass--metallicity evolution extracted from our sample will be conducted in a following paper.

\section{Data Availability}

The GAMA panchromatic photometry and redshifts are available to any member of the public via a collaboration request.\footnote{\url{http://www.gama- survey.org/collaborate/}} The photometric data come from the \texttt{GAMAKidsVikingFIRv01} Data Management Unit (DMU), the redshifts come from the \texttt{SpecObjv27} DMU, and the visual morphologies used in this paper come from the \texttt{VisualMorphologyv03} DMU.


\section{Acknowledgements}

The authors thank the anonymous referee, whose suggestions improved the paper. 

GAMA is a joint European-Australasian project based around a spectroscopic campaign using the Anglo-Australian Telescope. The GAMA input catalogue is based on data taken from the Sloan Digital Sky Survey and the UKIRT Infrared Deep Sky Survey. Complementary imaging of the GAMA regions is being obtained by a number of independent survey programmes including GALEX MIS, VST KiDS, VISTA VIKING, WISE, Herschel-ATLAS, GMRT and ASKAP providing UV to radio coverage. GAMA is funded by the STFC (UK), the ARC (Australia), the AAO, and the participating institutions. The GAMA website is \url{http://www.gama-survey.org/}.

SB and SPD acknowledge support by the \textit{Australian Research Council}'s funding scheme DP180103740. 
SB thanks the following individuals for reading the paper and providing feedback: Malcolm Bremer, Abdolhosein Hashemizadeh, Benne Holwerda, \'{A}ngel L\'{o}pez-S\'{a}nchez, and Lingyu Wang. 

This work was supported by resources provided by the \textit{Pawsey Supercomputing Centre} with funding from the \textit{Australian Government} and the \textit{Government of Western Australia}.

We have used \textsc{R} \citep{R} and \textsc{python} for our data analysis, and acknowledge the use of \textsc{Matplotlib} \citep{Hunter07} for the generation of plots in this paper. This research made use of \textsc{Astropy},\footnote{\url{http://www.astropy.org}} a community-developed core \textsc{python} package for astronomy \citep{Astropy13, Astropy18}.

\bibliographystyle{mnras}
\setlength{\bibsep}{0.0pt}
\bibliography{BibLibrary}

\appendix

\section{Comparison to Semi-Analytic Models and Simulations}
\label{sec:ModelComparison}

\begin{figure*}
	\centering
	\includegraphics[width=180mm]{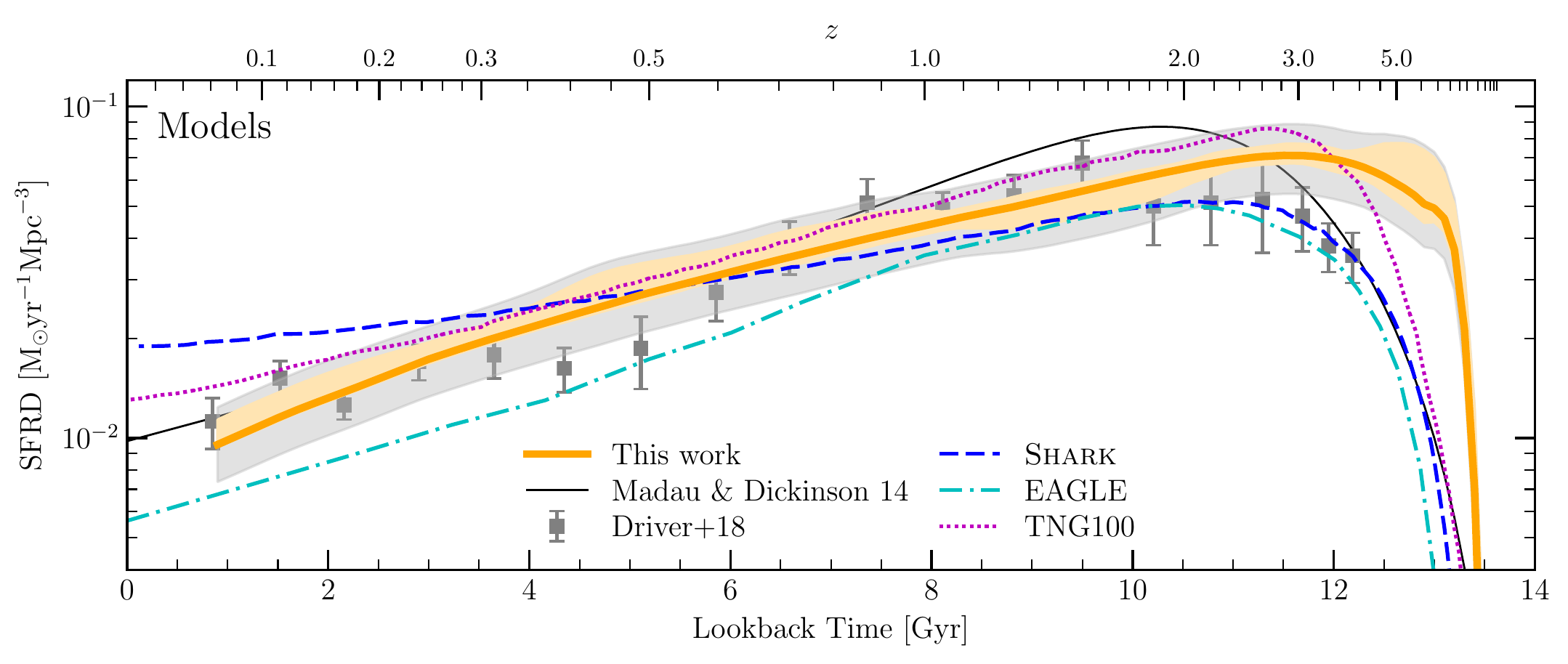}
	\caption{The cosmic star formation history from \textsc{ProSpect} (shown in orange), compared with observations and other theoretically determined star formation histories.  }
	\label{fig:cSFRComparisonTheory}
\end{figure*}

\begin{figure*}
	\centering
	\includegraphics[width=180mm]{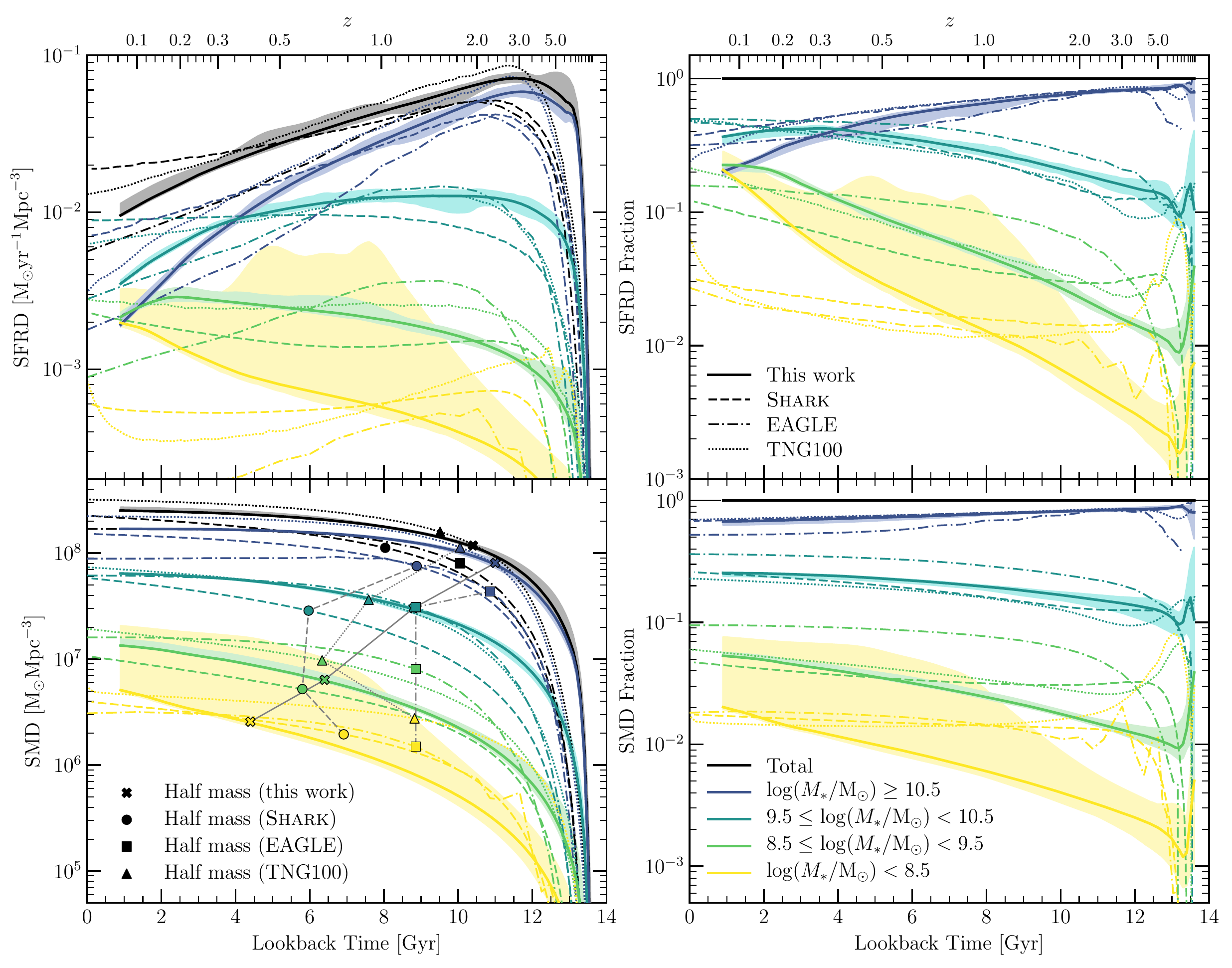}
	\caption{Right panel: the contribution of various stellar mass bins to the CSFH (top) and SMD (bottom), for this work as well as for the semi-analytic model \textsc{Shark} (dashed lines) and the simulations \textsc{EAGLE} (dashed-dotted lines) and TNG100 (dotted lines). In the bottom panel, the markers indicate the epoch at which each contribution has reached half of its final stellar mass (crosses for this work, circles for \textsc{Shark}, squares for \textsc{EAGLE}, and triangles for TNG100), connected by grey lines across the stellar mass bins. 
		Left panel: the fractional contribution of different stellar mass bins to the CSFH (top panel) and the SMD (bottom panel), as compared with \textsc{Shark},  \textsc{EAGLE}, and TNG100.  }
	\label{fig:SFRSMDFractions}
\end{figure*}

We compare our derived CSFH with that of the semi-analytic model (SAM) \textsc{Shark} \citep{Lagos18} and the cosmological, hydrodynamical simulations \textsc{EAGLE}\footnote{Data publicly available at \url{http://icc.dur.ac.uk/Eagle/database.php} \citep{McAlpine16}.} \citep{Schaye15} and IllustrisTNG\footnote{Data publicly available at \url{https://www.tng-project.org/data/} \citep{Nelson19}.} \citep[specifically the TNG100 box;][]{Pillepich18, Nelson18} in Fig. \ref{fig:cSFRComparisonTheory}. 
There are noticeable differences between them. While the CSFH curves for \textsc{Shark} and \textsc{EAGLE} are very similar in the early Universe, they begin to diverge at a lookback time of $\sim$ 10 Gyr. The CSFH from \textsc{Shark} is higher than the observationally derived CSFH at recent times, whereas the \textsc{EAGLE} CSFH is lower at recent times \citep[as also shown by][]{Furlong15}. 
The CSFH as given by TNG100 is higher than both \textsc{Shark} and \textsc{EAGLE} in the early Universe, but at recent times it is between the other two, roughly consistent with observed SFRD values. 

We identify how the SFRD and SMD curves for galaxies in different stellar mass bins compare to those derived for GAMA using \textsc{ProSpect} in Fig. \ref{fig:SFRSMDFractions}. 
The left panel of Fig. \ref{fig:SFRSMDFractions} is structurally the same as that of Fig. \ref{fig:SFD-SMD_Mass}, with the inclusion of equivalent results from each of the models for comparison. 
It is interesting to note that the systematic over/under-estimation of the CSFH by \textsc{Shark} and \textsc{EAGLE} can largely be explained by the star formation activity of the most massive galaxies alone. When analysing the SFRD for only the $\log(M_*/\rm{M}_{\odot}) \geq 10.5$ bin, the two curves for \textsc{Shark} and \textsc{EAGLE} start to diverge at a lookback time of $\sim$ 10 Gyr (just like the total CSFH curves). 
These curves are also systematically above and below the \textsc{ProSpect}-derived curves in this stellar mass bin, respectively. 
Additionally, the SFRD in the highest-mass bin for TNG100 is systematically greater than \textsc{ProSpect} at almost all epochs, especially for the last 4 Gyr.
This similarly reflects the differences between the TNG100 CSFH and that derived in this work. 
This emphasizes that, unsurprisingly, the shape of the CSFH is predominantly governed by the contribution of the most massive present-day galaxies. 
This also means that in using the CSFH to constrain simulations one may be inadvertently ignoring galaxies below the knee of the stellar mass function, suggesting that a combination of constraints that are sensitive to high- and low-mass galaxies are required.

The SFRD curves for the other mass bins vary significantly, both in normalisation and shape. The \textsc{ProSpect}-derived SFRD of the $9.5 \leq \log(M_*/\rm{M}_{\odot}) < 10.5$ bin remains roughly constant between lookback times of 12 and 6 Gyrs, at which time the SFRD starts to fall. Both \textsc{EAGLE} and TNG100 follow this trend broadly in this bin, although the SFRD drop-off is lower in TNG100. In \textsc{Shark}, however, the SFRD continues to rise incrementally until the present day. 
This is not surprising, as \citet{Bravo20} showed that the transition from predominantly star-forming to predominantly passive galaxies in \textsc{Shark} happens at slightly higher stellar masses than suggested by observations.

Similar discrepancies exist in the $8.5 \leq \log(M_*/\rm{M}_{\odot}) < 9.5$ bin. The \textsc{ProSpect}-derived SFRD is constantly rising up until 2 Gyr ago, at which time there is a turnover in the SFRD. The shape of the \textsc{EAGLE} SFRD in this stellar mass bin is most discrepant, peaking at a lookback time of $\sim$ 9 Gyr, and declining to the present day (a shape very similar to that of the total CSFH). The \textsc{Shark} SFRD in this bin has an early peak followed by a brief preiod of decline, but then rises slowly from a lookback time of 6 Gyr to the present day. TNG100, on the other hand, displays a trend very similar to that derived in this work, albeit without the drop in SFRD at recent times. 

While we include the results in the lowest-mass bin for completeness, we highlight that they should be treated with great caution. The results from this work using \textsc{ProSpect} are potentially biased by the mass incompleteness correction applied. In the results for \textsc{Shark}, this bin has a lower stellar mass limit of $\log(M_*/\rm{M}_{\odot}) =7$, whereas this limit for \textsc{EAGLE} is $\log(M_*/\rm{M}_{\odot}) =8$.  In the TNG100 lowest-mass bin, we have included the stellar mass and SFH for every stellar particle \emph{not} associated with a $\log(M_*/\rm{M}_{\odot})>8.5$ subhalo. 
Limitations in the simulations' resolution means any comparison in the lowest-mass bin may be limited.

The bottom-left panel of Fig. \ref{fig:SFRSMDFractions} shows the stellar mass build-up in each bin. We use different symbols to show the epoch at which each curve reaches half of its final stellar mass. Most interestingly, the downsizing trend we recover and discussed in Sec. \ref{sec:StellarMassTrends} is significantly weaker in \textsc{Shark} (circles) and \textsc{EAGLE} (squares). In TNG100 (triangles), this trend exists in the three highest stellar mass bins more strongly than in the two other models. 

Accounting for overall differences between the CSFH derived by the models, we show the fractional contributions of each stellar mass bin to the CSFH and SMD in the right panel of Fig. \ref{fig:SFRSMDFractions}. By presenting the data in this way, we effectively normalise differences between the total CSFH for each case. Arguably, this provides a fairer comparison between the models, especially in the early Universe where there are significant differences between the CSFH curves. This comparison shows that the fractional contributions of star formation and stellar mass are similar for each of the compared models in the two most massive stellar mass bins, but less so below $10^{9.5}\,\rm{M}_{\odot}$. 

The comparison here offers new avenues to understand the types of galaxies that determine the predicted CSFH and SMD. It is clear that the simulations shown here display various degrees of zeal in their star formation activity that depend on the cosmic epoch and stellar mass analysed. 
Even at $z\sim0$, where the \textsc{ProSpect} SFR outputs are most robust to modelling assumptions, all simulations disagree with the observations. This is especially true for the most massive galaxies, where while \textsc{EAGLE} is close, both \textsc{Shark} and TNG100 overpredict the SFRD. 
However, one has to be cautious as there are several systematic effects that we have not considered. Ideally we would use \textsc{ProSpect} to fit the predicted SEDs of simulated galaxies and derive a SFH and ZH for each simulated galaxy to compare with the GAMA ones reported here. This is however, beyond the scope of this paper and we leave it for future work. 

\section{Alternative metallicity implementation}
\label{sec:LinearMetallicity}

\begin{figure*}
	\centering
	\includegraphics[width=180mm]{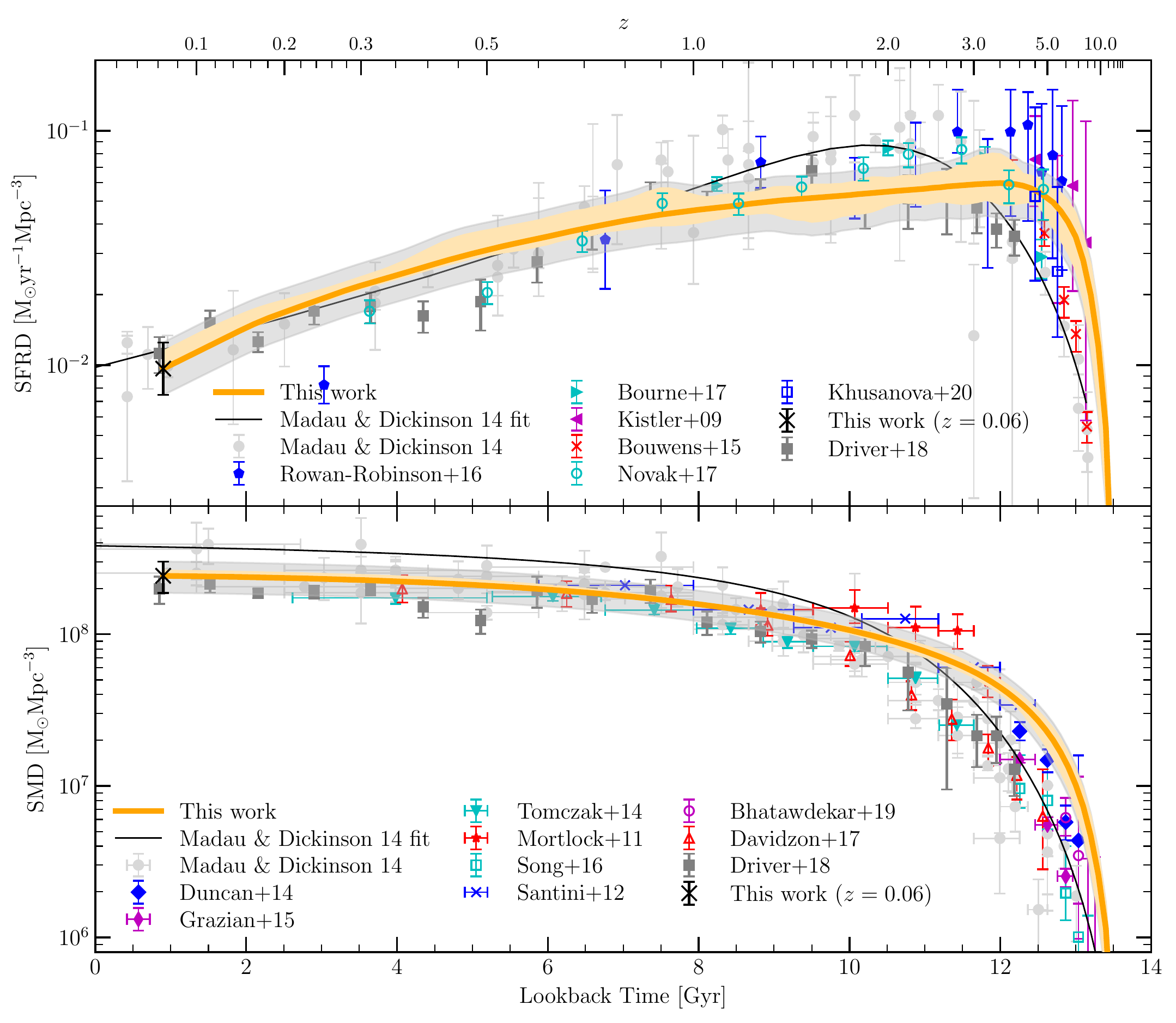}
	\caption{The cosmic star formation history (top panel) and stellar mass density (bottom panel) from \textsc{ProSpect} (shown in orange), using the \texttt{Zfunc\_massmap\_lin} implementation of metallicity evolution.    }
	\label{fig:Lin}
\end{figure*}

As outlined by \citet{Robotham20}, there are two methods of implementing a reasonable evolution of metallicity for individual galaxies within \textsc{ProSpect}. One of these is the closed-box metallicity evolution, as given by the \texttt{Zfunc\_massmap\_box} function, which we have used in the main body of this paper. An alternative to this is the so-called ``linear" metallicity evolution, as given by the \texttt{Zfunc\_massmap\_lin} function, which linearly maps the build-up of stellar mass onto the build-up of metallicity. The main difference between this model and the closed-box implementation, in effect, is that the yield is allowed to evolve with time. As a result, the late-time enrichment for galaxies with the \texttt{Zfunc\_massmap\_lin} model is slightly lower than that for the \texttt{Zfunc\_massmap\_box} model. This difference is highlighted in fig. 14 of \citet{Robotham20}. 

Fig. \ref{fig:Lin} is equivalent to Fig. \ref{fig:cSFRComparisonLiterature}, but now shows the SFRD and cSMD derived using the linear metallicity evolution. While the results are still broadly consistent with the literature (and the differences resulting from this implementation are subtle), we see that the resulting SFRD is now flatter than before, with the peak shifting from $\sim$ 12 Gyr ago, to a very broad peak between $\sim$ 8--12 Gyr ago. In contrast, the SMD is in much closer agreement with observations, especially in the first $\sim$ 2 Gyr of the Universe. 
The SFRD value at $z=0.06$ here is measured to be $(9.7^{+2.8}_{-2.2})\times 10^{-3}\, \rm{M}_{\odot}\,\rm{yr}^{-1}\,\rm{Mpc}^{-3}$, and the equivalent measurement for the SMD is $(2.43^{+0.59}_{-0.55})\times10^8\, \rm{M}_{\odot}\,\rm{Mpc}^{-3}$. These values are shown in Fig. \ref{fig:Lin} as black crosses. Note that they are consistent with the values derived using the \texttt{Zfunc\_massmap\_box} implementation. 
The literature values of the SFRD and SMD, which have been measured using distinct techniques, are not consistent with each other. We note that, given slightly different assumptions of a reasonable metallicity history for individual galaxies, we derive a result that is either in better agreement with the observed SFRD values, or observed SMD values. 

\section{Tabulated data}


\begin{table*}
	\centering
	\caption[SFRD subsets]{SFRD subsets plotted in Figures \ref{fig:SFD-SMD_Mass} and \ref{fig:SFD-SMD_Morph}.  The boundaries for each of the stellar mass bins are: Mass 1 = $\log(M_*/\rm{M}_{\odot}) \geq 10.5$; Mass 2 = $9.5 \leq \log(M_*/\rm{M}_{\odot}) < 10.5$; Mass 3 = $8.5 \leq \log(M_*/\rm{M}_{\odot}) < 9.5$; and Mass 4 = $8.5 \leq \log(M_*/\rm{M}_{\odot})$. Full table available \href{https://github.com/SabineBellstedt/Bellstedt2020--SupplementaryData}{online}. }
	\label{tab:SFRDvalues}
	\begin{tabular}{@{}cc | ccccccccc}
		\hline
		Lookback Time& $z$ & cSFRD & \texttt{E} & \texttt{S0-Sa} & \texttt{Sab-Scd} & \texttt{Sd-Irr} & Mass 1 & Mass 2 & Mass 3 & Mass 4 \\
		Gyr && \multicolumn{9}{c}{$\rm{M}_{\odot}\rm{yr}^{-1} \rm{Mpc}^{-3}$}   \\
		\hline
		\hline
1.0 & 0.07 & 0.0099 & 0.0005 & 0.0015 & 0.0037 & 0.0030 & 0.0020 & 0.0037 & 0.0022 & 0.0020 \\ 
1.2 & 0.09 & 0.0105 & 0.0006 & 0.0017 & 0.0040 & 0.0031 & 0.0023 & 0.0040 & 0.0024 & 0.0019 \\ 
1.4 & 0.10 & 0.0112 & 0.0007 & 0.0020 & 0.0043 & 0.0031 & 0.0025 & 0.0043 & 0.0025 & 0.0018 \\ 
1.6 & 0.12 & 0.0119 & 0.0007 & 0.0022 & 0.0046 & 0.0031 & 0.0028 & 0.0047 & 0.0026 & 0.0018 \\ 
1.8 & 0.14 & 0.0126 & 0.0008 & 0.0024 & 0.0050 & 0.0031 & 0.0031 & 0.0050 & 0.0027 & 0.0017 \\ 
2.0 & 0.15 & 0.0133 & 0.0009 & 0.0026 & 0.0053 & 0.0031 & 0.0035 & 0.0053 & 0.0028 & 0.0016 \\ 
... & &  & && & & & & &  \\ 
\hline
	\end{tabular}
\end{table*}

\begin{table*}
	\centering
	\caption[SMD subsets]{SMD subsets plotted in Figures \ref{fig:SFD-SMD_Mass} and \ref{fig:SFD-SMD_Morph}. The boundaries for each of the stellar mass bins are: Mass 1 = $\log(M_*/\rm{M}_{\odot}) \geq 10.5$; Mass 2 = $9.5 \leq \log(M_*/\rm{M}_{\odot}) < 10.5$; Mass 3 = $8.5 \leq \log(M_*/\rm{M}_{\odot}) < 9.5$; and Mass 4 = $8.5 \leq \log(M_*/\rm{M}_{\odot})$. Full table available \href{https://github.com/SabineBellstedt/Bellstedt2020--SupplementaryData}{online}. }
	\label{tab:SMDvalues}
	\begin{tabular}{@{}cc | ccccccccc}
		\hline
		Lookback Time &$z$& cSMD & \texttt{E} & \texttt{S0-Sa} & \texttt{Sab-Scd} & \texttt{Sd-Irr}  & Mass 1 & Mass 2 & Mass 3 & Mass 4 \\
		Gyr && \multicolumn{9}{c}{$\log(\rm{M}_{\odot}\rm{Mpc}^{-3})$}   \\
		\hline
		\hline
1.0 & 0.07 & 8.4017 & 7.9064 & 7.9743 & 7.7012 & 7.1909 & 8.2292 & 7.8074 & 7.1258 & 6.6995 \\ 
1.2 & 0.09 & 8.4005 & 7.9064 & 7.9741 & 7.6988 & 7.1831 & 8.2290 & 7.8056 & 7.1195 & 6.6824 \\ 
1.4 & 0.10 & 8.3991 & 7.9063 & 7.9737 & 7.6961 & 7.1750 & 8.2288 & 7.8036 & 7.1126 & 6.6649 \\ 
1.6 & 0.12 & 8.3977 & 7.9063 & 7.9733 & 7.6932 & 7.1664 & 8.2286 & 7.8014 & 7.1051 & 6.6468 \\ 
1.8 & 0.14 & 8.3961 & 7.9062 & 7.9729 & 7.6899 & 7.1569 & 8.2283 & 7.7990 & 7.0970 & 6.6249 \\ 
2.0 & 0.15 & 8.3944 & 7.9062 & 7.9724 & 7.6861 & 7.1472 & 8.2281 & 7.7962 & 7.0878 & 6.6039 \\ 
... & &  & && & & & & &  \\ 
\hline
	\end{tabular}
\end{table*}

\begin{table}
	\centering
	\caption[Metal Density subsets]{Cosmic metal mass density values plotted in Fig. \ref{fig:CosmicMetalDensity}. Full table available \href{https://github.com/SabineBellstedt/Bellstedt2020--SupplementaryData}{online}. }
	\label{tab:MetalDensityvalues}
	\begin{tabular}{@{}cc | ccc}
		\hline
		Lookback Time& $z$ & Total & Stars & Gas  \\
		Gyr && \multicolumn{3}{c}{$\log(\rm{M}_{\odot}\rm{Mpc}^{-3})$}   \\
		\hline
1.0 & 0.07 & 6.7829 & 6.5690 & 6.3718 \\ 
1.2 & 0.09 & 6.7813 & 6.5665 & 6.3715 \\ 
1.4 & 0.10 & 6.7795 & 6.5639 & 6.3711 \\ 
1.6 & 0.12 & 6.7776 & 6.5611 & 6.3707 \\ 
1.8 & 0.14 & 6.7756 & 6.5581 & 6.3702 \\ 
2.0 & 0.15 & 6.7735 & 6.5550 & 6.3696 \\ 
... & &  & & \\ 
		\hline

	\end{tabular}
\end{table}

\label{lastpage} 
\end{document}